\documentclass[prb,amsfonts,amssymb,floats,twocolumn,superscriptaddress,aps]{revtex4-1}
\usepackage{amsmath}
\usepackage{float}
\usepackage[caption = false]{subfig}
\usepackage{color}
\usepackage{braket}
\usepackage{bm}
\usepackage{dsfont}
\usepackage{adjustbox}
\usepackage{graphicx}
\usepackage{tikz}
\usepackage{placeins}
\usepackage{bbm}

\allowdisplaybreaks

\newcommand{\jp}{J_\perp} 
\newcommand{\jpk}{J_\perp / K} 

\newcommand{\sac}{\mbox{$\sigma$AC}}
\newcommand{\sbac}{\mbox{$\bar{\sigma}$AC}}

\newcommand{\ii}{\mathrm{i}} 
\newcommand{\eu}{\mathrm{e}} 

\newcommand{\kk}{\vec{k}}

\newcommand{\pdksli}{KSL}
\newcommand{\pdkslii}{FLUX}
\newcommand{\pddim}{DIM}
\newcommand{\pdcross}{DIM'}
\newcommand{\pdchains}{MAC}
\newcommand{\pdmacaf}{MAC-AF}
\newcommand{\pdmacsl}{MAC-L}

\newcommand{\Ztwo}{\mathbb{Z}_2}

\newcommand{\SUtwo}{\mathrm{SU(2)}}
\newcommand{\SOfour}{\mathrm{SO(4)}}

\newcommand{\bvec}[1]{\bm{#1}}
\newcommand{\bmat}[1]{\bm{#1}}

\newcommand{\anb}{b^{\phantom\dagger}}
\newcommand{\crb}{b^\dagger}
\newcommand{\bn}{{\boldsymbol{n}}}
\newcommand{\bi}{{\boldsymbol{i}}}

\newcommand{\sqtwo}{{
\begin{adjustbox}{minipage=0.4cm, padding=0pt -0.8cm 0pt 0pt}
\begin{tikzpicture}[scale=0.2]
\draw (0,0) -- (1,0);
\draw (0,0) -- (0,1);
\draw (0,1) -- (1,1);
\draw (1,0) -- (1,1);
\draw (1,0) -- (2,0);
\draw (2,0) -- (2,1);
\draw (1,1) -- (2,1);
\draw (0,1) -- (0,2);
\draw (0,2) -- (1,2);
\draw (1,1) -- (1,2);
\draw (1,2) -- (2,2);
\draw (2,1) -- (2,2);
\draw [fill=black, fill opacity=0.4] (0,1) rectangle (1,2);
\draw [fill=black, fill opacity=0.4] (1,0) rectangle (2,1);
\end{tikzpicture}
\end{adjustbox}}
}

\newcommand{\sqthree}{{
\begin{adjustbox}{minipage=0.4cm, padding=0pt -0.8cm 0pt 0pt}
\begin{tikzpicture}[scale=0.2]
\draw (0,0) -- (1,0);
\draw (0,0) -- (0,1);
\draw (0,1) -- (1,1);
\draw (1,0) -- (1,1);
\draw (1,0) -- (2,0);
\draw (2,0) -- (2,1);
\draw (1,1) -- (2,1);
\draw (0,1) -- (0,2);
\draw (0,2) -- (1,2);
\draw (1,1) -- (1,2);
\draw (1,2) -- (2,2);
\draw (2,1) -- (2,2);
\draw [fill=black, fill opacity=0.4] (0,0) rectangle (1,1);
\draw [fill=black, fill opacity=0.4] (1,1) rectangle (2,2);
\end{tikzpicture}
\end{adjustbox}}
}

\newcommand{\sqfour}{{
\begin{adjustbox}{minipage=0.4cm, padding=0pt -0.8cm 0pt 0pt}
\begin{tikzpicture}[scale=0.2]
\draw (0,0) -- (1,0);
\draw (0,0) -- (0,1);
\draw (0,1) -- (1,1);
\draw (1,0) -- (1,1);
\draw (1,0) -- (2,0);
\draw (2,0) -- (2,1);
\draw (1,1) -- (2,1);
\draw (0,1) -- (0,2);
\draw (0,2) -- (1,2);
\draw (1,1) -- (1,2);
\draw (1,2) -- (2,2);
\draw (2,1) -- (2,2);
\draw [fill=black, fill opacity=0.4] (0,0) rectangle (2,2);
\end{tikzpicture}
\end{adjustbox}}
}

\begin{document}

\title[]
{
Bilayer Kitaev models: Phase diagrams and novel phases
}
\author{Urban F. P. Seifert}
\affiliation{Institut f\"ur Theoretische Physik,
Technische Universit\"at Dresden, 01062 Dresden, Germany}
\author{Julian Gritsch}
\affiliation{Institut f\"{u}r Theoretische Physik I, Universit\"at Erlangen-N\"urnberg, 91058 Erlangen, Germany}
\author{Erik Wagner}
\affiliation{Institut f\"{u}r Theoretische Physik, Technische Universit\"at Braunschweig, 38106 Braunschweig, Germany}
\author{Darshan G. Joshi}
\affiliation{Max-Planck-Institut f\"ur Festk\"orperforschung, 70569 Stuttgart, Germany}
\author{Wolfram Brenig}
\affiliation{Institut f\"{u}r Theoretische Physik, Technische Universit\"at Braunschweig, 38106 Braunschweig, Germany}
\affiliation{Institut f\"ur Theoretische Physik,
Technische Universit\"at Dresden, 01062 Dresden, Germany}
\author{Matthias Vojta}
\affiliation{Institut f\"ur Theoretische Physik,
Technische Universit\"at Dresden, 01062 Dresden, Germany}
\author{Kai P. Schmidt}
\affiliation{Institut f\"{u}r Theoretische Physik I, Universit\"at Erlangen-N\"urnberg, 91058 Erlangen, Germany}


\date{\today}

\begin{abstract}
Kitaev's honeycomb-lattice spin-$1/2$ model has become a paradigmatic example for {$\Ztwo$} quantum spin liquids, both gapped and gapless. Here we study the fate of these spin-liquid phases in differently stacked bilayer versions of the Kitaev model. Increasing the ratio between the inter-layer Heisenberg coupling $\jp$ and the intra-layer Kitaev couplings $K^{x,y,z}$ destroys the topological spin liquid in favor of a paramagnetic dimer phase.
We study phase diagrams as a function of $\jp/K$ and Kitaev coupling anisotropies using Majorana-fermion mean-field theory, and we employ different expansion techniques in the limits of small and large $\jp/K$.
For strongly anisotropic Kitaev couplings, we derive effective models for the different layer stackings which we use to discuss the quantum phase transition out of the Kitaev phase.
We find that the phase diagrams depend sensitively on the nature of the stacking and anisotropy strength. While in some stackings and at strong anisotropies we find a single transition between the Kitaev and dimer phases, other stackings are more involved:
Most importantly, we prove the existence of two novel macro-spin phases which can be understood in terms of Ising chains which can be either coupled ferromagnetically, or remain degenerate, thus realizing a classical spin liquid. In addition, our results suggest the existence of a flux phase with spontaneous inter-layer coherence.
We discuss prospects for experimental realizations.
\end{abstract}

\pacs{}

\maketitle


\section{Introduction}
\label{sec:intro}

\begin{figure*}[!tb]
\includegraphics[width=0.95\textwidth,clip]{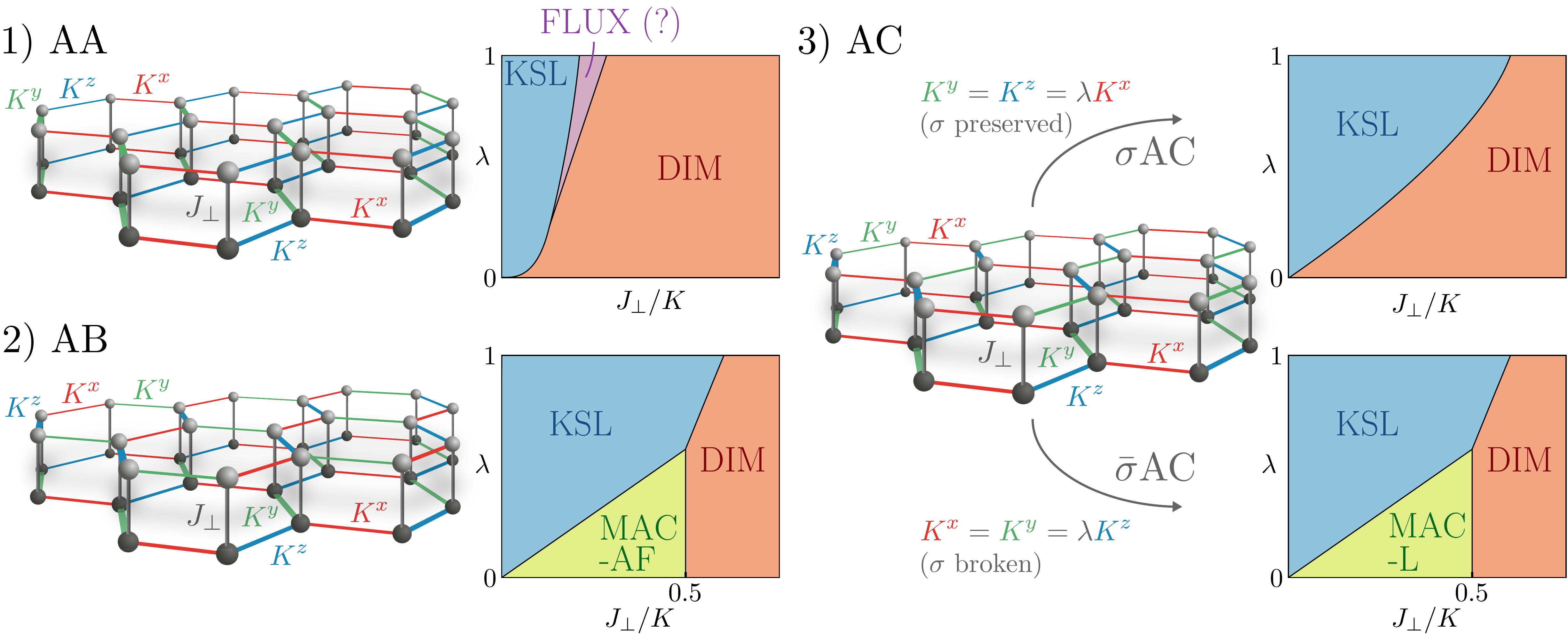}
\caption{Illustration of different stackings AA, AB and AC (with two inequivalent choices of anisotropy), and schematic phase diagrams. The critical $\jpk$ for the transition from the \pdmacaf{} and \pdmacsl{} phases to the \pddim{} phase is obtained exactly at $\lambda=0$ to be $\jp = K / 2$ with $K=\mathrm{max}(K^x,K^y,K^z)$. As explained in Sec.~\ref{sec:dimer_stab}, the \pddim{} phase in the AA stacking is expected to be of greater stability compared to the AB- and AC-stacked models because triplons in the AA stacking are fully localized. For various labels and further details we refer the reader to the text. \label{fig:header}
}
\end{figure*}

Quantum spin liquids\cite{anderson73,savary_rop17,kanoda_rmp17} constitute a fascinating class of many-body phases which have attracted tremendous attention over the past decades: On the one hand, they feature properties like fractionalization, topological order, and long-range entanglement, all of fundamental interest in the context of understanding and classifying phases beyond Landau's paradigm of symmetries and spontaneous symmetry breaking. On the other hand, quantum spin liquids may hold the key to understanding unsolved puzzles in the physics of cuprate superconductors,\cite{leermp} and their excitations have been discussed as elements for topological quantum computation.\cite{kitaev03,kitaev06}

Kitaev's honeycomb-lattice model\cite{kitaev06} is a particular spin model realizing a quantum spin liquid with emergent {$\Ztwo$} gauge structure. It is a rare example of an exactly solvable spin model in two space dimensions (2D), thanks to an infinite number of conserved quantities, which has allowed the community to obtain a large number of exact or quasi-exact results, including dynamical spin correlations\cite{knolle} as well as thermodynamic and transport properties.\cite{nasu15,nasu17,brenig17} Generalizations to other lattices, preserving the exact solubility, have been proposed in both 2D and 3D.\cite{yao07,dusuel08,baskaran09,kiv1,mandal09,kamfor10,trebst16} Moreover, the Kitaev model has been used as a controlled starting point for investigations beyond integrability, for instance targeting metallic and superconducting phases of systems with highly anisotropic magnetic interactions.\cite{ykv12,halasz14,smv17,mei12}

In this paper, we consider different bilayer versions of the Kitaev model, with antiferromagnetic Heisenberg coupling between the layers. The motivation is twofold:
(i) Given that the two limits -- a {$\Ztwo$} spin liquid and a featureless dimer paramagnet -- are phases without spontaneously broken symmetries, a key question is whether they are separated by a single (topological) quantum phase transition, or whether additional phases -- with or without symmetry breaking -- intervene.
(ii) For transitions without symmetry breaking it is interesting to determine their characteristics. For instance, a continuous transition out of a {$\Ztwo$} topological phase, i.e., a spinon confinement transition, is expected to be driven by the condensation of visons and described by a {$\Ztwo$} gauge theory. \cite{read91,wen91,ms01,ropfru,ssbook}
This continuous quantum phase transition has indeed been found in perturbing the anisotropic limit of the honeycomb Kitaev model.\cite{Trebst07,Hamma08,Vidal09_1,Tupitsyn10,Dusuel11}

We attack the problem from different directions: We employ a Majorana-based mean-field theory, which enables us to determine phase diagrams covering the entire parameter space and moreover becomes exact in the isolated-layer limit. In addition, we use bond-operator and series-expansion techniques to describe the dimer phase at strong interlayer coupling and its breakdown. Finally, we construct effective models in the limit of strongly anisotropic Kitaev coupling (i.e. the toric-code limit\cite{kitaev03}), which we use to study the phases and transitions in this limit.

\subsection*{Summary of results}
The main results can be summarized as follows:
Different stackings of the Kitaev $x,y,z$ bonds, yielding different symmetry properties, produce significantly different phase diagrams, as summarized in Fig.~\ref{fig:header}.
These differences are particularly pronounced at strong anisotropies of the Kitaev couplings.

In the following, we denote the Kitaev couplings by $K$ (in the isotropic case), while the $J_\perp$ is the antiferromagnetic interlayer coupling. We introduce an anisotropy for the Kitaev layers by rescaling two of the three Kitaev couplings as $\lambda K$, where $0 \leq \lambda \leq 1$, such that $\lambda = 0$ for a single-layer Kitaev model yields decoupled (in-plane) dimers, while $\lambda = 1$ corresponds to the isotropic Kitaev model (see Sec.~\ref{sec:model} and Fig.~\ref{fig:header} for a definition of the stackings and further notational details).

For the AB and \sbac{} stackings, novel macro-spin phases (MAC) appear.
The building blocks of MAC are emergent Ising macro-spins. Each chain, formed from the interlayer coupling and a strong Kitaev coupling, constitutes a macro-spin.
These chains can be mapped, in the fully anisotropic limit, to an Ising chain in a transverse field.
Given the exact solubility of this effective model, the phase diagrams thus become exact in the anisotropic limit $\lambda \to 0$. In particular, the transition from MAC to the trivial dimer phase (DIM) is located at $\jp/K = 0.5$.
Notably, in the \sbac{} stacking this collection of Ising macro-spins has a macroscopically large degeneracy even at finite inter-chain couplings (i.e. when going away from the anisotropic limit), thus realizing a classical spin liquid, while in the AB stacking the coupling of the macro-spins leads to antiferromagnetic long-range order.

In the AA and \sac{} stackings, the anisotropic limit $\lambda \to 0$ leads to decoupled in-plane dimers (inter-layer plaquettes), and only a single transition between the Kitaev spin liquid (KSL) and DIM occurs. We perturbatively derive effective models for the breakdown of KSL in the anisotropic limit and at small $\jp$ for both stackings.
We then perform a mapping to a dual Ising model (with higher-order plaquette-interactions) for the AA stacking, exploiting the fact that there are conserved quantities at finite $\jp$ (in contrast to the other stackings). In the dual effective model the transition from the topological KSL to the trivial dimer phase corresponds to a transition from a pseudo-spin-polarized state to a symmetry-broken phase. 
The analysis of the dual model shows that the KSL-DIM transition in the AA stacking lies in the (2+1)D Ising universality class, and the critical interlayer coupling scales as $\jp/K \propto \lambda^4$, where $\lambda$ parametrizes the anisotropy. We further argue that in the \sac{} stacking the transition can be expected to be of first order.

Finally, our mean-field results suggest the existence of a phase (dubbed FLUX) with spontaneous interlayer coherence. This phase masks the transition between KSL and DIM close to the isotropic point in the AA stacking and is akin to an exciton condensate phase.
We find that inversion symmetry is spontaneously broken in this phase, resulting in the spontaneous formation of $\pi$-fluxes of the $\Ztwo$ gauge field in interlayer plaquettes. Importantly, broken inversion symmetry allows the itinerant Majorana fermions to be gapped for all parameter regimes. This is to be contrasted to KSL, for which we argue that no single-Majorana hopping processes can occur, which also implies that the nodal points in the spectrum are protected.

\subsection*{Outline}

The remainder of the paper is organized as follows:
In Sec.~\ref{sec:model} we introduce the bilayer Kitaev models and discuss their symmetry properties.
Sec.~\ref{sec:mf} describes the Majorana mean-field theory and its results for the different stackings, in particular mean-field phase diagrams.
In Sec.~\ref{sec:series_expansion} we outline the series expansion techniques used.
Sec.~\ref{sec:series_expansion_dimer} presents the results for the one-triplon dispersion obtained through series expansion in the paramagnetic phase for various stackings and anisotropies.
In Sec.~\ref{sec:qpt_eff_mod_aa} we focus on the limit of strongly anisotropic Kitaev couplings, where controlled analytical progress can be made. In particular, we derive effective models for two different stackings which allow us to deduce properties of phases and phase transitions in the anisotropic limit.
The novel macro-spin phases are discussed in Sec.~\ref{sec:novel}, where we derive effective models and discuss both the antiferromagnetic state as well as the classical spin liquid.
In Sec.~\ref{sec:pert_ksl} we discuss the stability of the Kitaev spin liquid against a small inter-layer coupling, and the possibility of flux phase with spontaneous inter-layer coherence.
A discussion closes the paper.



\section{Model and stacking}
\label{sec:model}

We consider two stacked honeycomb layers, with spins $1/2$ on each lattice site denoted by $S_{mi}$ where $i$ is a site index in each layer and $m=1,2$ is the layer index, such that $i$ also labels inter-layer dimers of adjacent sites.
In our study, we assume that the two layers are stacked such that the sites of two layers are on top of each other, as opposed to, e.g., Bernal stackings.

\subsection{Hamiltonian}

The construction of the Kitaev model is based on distinguishing three sets of mutually parallel bonds on the honeycomb lattice; we will denote these sets by $1,2,3$. In a single-layer Kitaev model, each set is assigned to a spin component, $123\to xyz$, to form Ising bonds. For the bilayer model, we will use identical bond numbers for both layers, and define a layer Hamiltonian as follows:
\begin{align}
\label{eq:h1}
\mathcal{H}_m^{\alpha\beta\gamma} &=
-\sum_{\langle ij\rangle_1} K^\alpha S_{mi}^\alpha S_{mj}^\alpha
-\sum_{\langle ij\rangle_2} K^\beta S_{mi}^\beta S_{mj}^\beta \nonumber\\
&-\sum_{\langle ij\rangle_3} K^\gamma S_{mi}^\gamma S_{mj}^\gamma
\end{align}
where $\langle ij \rangle$ denotes nearest-neighbor sites.
We assume ferromagnetic Kitaev interactions, $K^\alpha > 0$, however there is a duality transformation which inverts all Kitaev couplings $K^\alpha \to - K^\alpha$ (cf. Sec. \ref{sec:modelsymm}), such that the results for ferromagnetic Kitaev couplings presented in this paper also hold for antiferromagnetic $K^\alpha$.
The interlayer coupling is of Heisenberg type with strength $\jp > 0$:
\begin{align}
\mathcal{H}_\perp &= \jp \sum_{i} \vec{S}_{1i} \cdot \vec{S}_{2i}\,.
\end{align}
%
%
%

For the bilayer model, different stackings of the bond flavors $xyz$ are possible, as illustrated in Fig.~\ref{fig:header}. The case with identical bonds in both layers, dubbed AA stacking in the following, is described by the Hamiltonian
\begin{align}
\label{eq:haa}
\mathcal{H}_{\rm AA} &= \mathcal{H}_1^{xyz} + \mathcal{H}_2^{xyz} + \mathcal{H}_\perp \,;
\end{align}
of course, simultaneous cyclic permutations of the bond flavors on \emph{both} lattices lead to equivalent models.
Using different flavor assignments on both layers, various additional distinct stackings are possible, such as
\begin{align}
\label{eq:hab}
\mathcal{H}_{\rm AB} &= \mathcal{H}_1^{xyz} + \mathcal{H}_2^{yzx} + \mathcal{H}_\perp \,, \\
\label{eq:hac}
\mathcal{H}_{\rm AC} &= \mathcal{H}_1^{xyz} + \mathcal{H}_2^{xzy} + \mathcal{H}_\perp \,.
\end{align}
These stackings lead to different symmetry properties of the full Hamiltonian, as we will discuss below.

We will consider the isotropic Kitaev models as well as the case of anisotropic couplings. As shown by Kitaev,\cite{kitaev06} increasing the anisotropy in a single-layer Kitaev model eventually gaps out the nodal points in the Majorana dispersion.
In the limit of strong anisotropy, the gapped phase can be mapped to Kitaev's toric code.\cite{kitaev03}

In the following, we parameterize the anisotropy in the AA and AB stacking as $K^x = K^y = \lambda K^z$ where $0 \leq \lambda \leq 1$, yielding one strong and two weak bonds.
For the AC stacking however, there are two inequivalent choices of anisotropy, depending if a mirror reflection $\sigma$ along the $x$-bonds (see Sec. \ref{sec:modelsymm} for an extended discussion of the symmetries of the model) is preserved under the anisotropy, with $K^y = K^z = \lambda K^x$, or broken, for instance by choosing $K^x = K^y = \lambda K^z$.
For notational convenience, we will call the former the \sac{} stacking, and use \sbac{} to refer to the latter case. Obviously, at $\lambda = 1$ these two notations refer to the same model.
We also introduce the notation $K = \mathrm{max}(K^x,K^y,K^z)$ to mark the largest of the Kitaev couplings.

It is clear that for dominant $\jp \gg K$ the two spins within an interlayer dimer form a spin-zero singlet independent of the stacking, such that the full system is a featureless quantum paramagnet. In the opposite limit $\jp \ll K$ the system consists of two weakly coupled {$\Ztwo$} spin liquids which can be expected to be stable.
An important difference between the stackings exists in the anisotropic limit $\lambda\to 0$, where each Kitaev layer consists of Ising-coupled dimers: $\jp$ couples these dimers \emph{either} to four-spin plaquettes (as is the case for the AA and \sac{} stackings)  \emph{or} to chains (AB, \sbac{}), and this difference turns out to be important, see Sec.~\ref{sec:qpt_eff_mod_aa} for more details.

\subsection{Symmetries and conservation laws}
\label{sec:modelsymm}

The bilayer model with AA stacking inherits all symmetries of the single-layer Kitaev model. In the isotropic case ($\lambda=1$) these can be generated from the following unitary operations:
(i) a $C_3$ lattice rotation combined with permuting the spin components $S^x \to S^y \to S^z \to S^x$,
(ii) a reflection symmetry $\sigma$ across an axis perpendicular to the $x$ bonds combined with the spin transformation \mbox{$S^x \to -S^x$}, $S^y \to -S^z$, $S^z \to -S^y$, and
(iii) an inversion of two spin components by a $\pi$-rotation around the $x$-axis, \mbox{$C_x^\ast$: $(S^x, S^y, S^z) \to (S^x, -S^y, -S^z)$}, and similarly for $C_y^\ast$, $C_z^\ast$.
In addition, the AA-stacked model is trivially symmetric under
(iv) layer exchange: $\vec{S}_{1i} \leftrightarrow \vec{S}_{2i}$.

The $\pi$-spin-rotation symmetry (iii) is a local operation and is thus also preserved for all variations of the stackings.
Notably, this particular local spin rotation symmetry implies that the models considered here are symmetric under the inversion of the Kitaev couplings $K^\alpha \to - K^\alpha$, as we can find an operation $U$ under which the ferromagnetic Kitaev model is mapped to the antiferromagnetic Kitaev model,\cite{fn:symmop} and this symmetry operation can be chosen to be identical in each layer, such that $U$ leaves the interlayer-coupling $\jp \vec S_{1i} \cdot \vec S_{2i}$ invariant.

For the AB stacking, the $C_3$ rotation symmetry is preserved, but the reflection symmetry (ii) is absent, and the layer exchange (iv) is only a symmetry if followed by a $C_3$ rotation.

Finally, in the case of AC stacking, the $C_3$ rotation is not a symmetry, while there is a reflection symmetry across the bond with the same interaction in both layers, i.e. the $x$-bond in the model defined in Eqn. \eqref{eq:hac}. Analogous to the AB stacked model, layer exchange is a symmetry if combined with the reflection operation $\sigma$.

Introducing a finite anisotropy ($\lambda < 1$), all symmetries [except (iii)] detailed above are spoiled, with the exception that in the AA- and \sac{}-stacked models, a reflection symmetry across the strong bond is retained.

While the single-layer Kitaev model is characterized by the conservation of Ising fluxes
\begin{equation}
\label{eq:wp}
 \hat{W}^p = S_1^x S_2^y S_3^z S_4^x S_5^y S_6^z
\end{equation}
for sites $1,\ldots,6$ along each individual plaquette, this conservation law is spoiled by the interlayer coupling.
However, for the AA stacking, the product of fluxes in intra-layer pairs of plaquettes, \mbox{$\hat{\Omega}_p\equiv\hat{W}_1^p \;\hat{W}_2^p$}, is still conserved. This implies a thermodynamically large number of conserved quantities, but cannot be obviously used to solve the model exactly.
In contrast, for both AB and AC stackings, there are no such conserved fluxes.


\section{Majorana mean-field theory}
\label{sec:mf}

In the following section, we employ a Majorana-based mean-field theory in order to map out the full phase diagram of the model. The advantage of our approach is that the mean-field theory is \emph{exact} in the limit $\jp = 0$, i.e. reproduces the Kitaev spin liquid physics for the two decoupled layers.


\subsection{Majorana representation}
\label{subsec:majrep}

The Kitaev honeycomb model defined on each layer \eqref{eq:h1} can be solved exactly\cite{kitaev06} by introducing four Majorana fermions $\chi^\mu$ with the anticommutation relations $\{ \chi^\mu, \chi^\nu\} = \delta^{\mu \nu}$.
The spin representation $S^\alpha_\mathrm{K} = \ii \chi^0 \chi^\alpha$ reproduces the $\SUtwo$ spin algebra as long as the (gauge) constraint $D \equiv 4 \chi^0 \chi^1 \chi^2 \chi^3 = 1$ is satisfied.

In order to elucidate pecularities of Kitaev's spin representation, we make the connection to more conventional slave-fermion approaches which decompose the spin as $S^\alpha = f_{\sigma}^\dagger \tau^\alpha_{\sigma \sigma'} f_\sigma / 2$ with two canonical fermions $f_\uparrow, f_\downarrow$. Mapping above expression to Majorana fermions with $f_\uparrow = (\chi^0 + \ii \chi^3)/\sqrt{2}$ and $f_\downarrow = (\ii \chi^1 - \chi^2) / \sqrt{2}$ then yields a different Majorana spin representation that uses all four Majorana fermions per site,
\begin{equation} \label{eq:spinrep}
  S^\alpha = \frac{\ii}{2} \left( \chi^0 \chi^\alpha - \frac{\ii}{2} \epsilon^{\alpha \beta \gamma} \chi^\beta \chi^\gamma \right) = \frac{\ii}{4} \bvec{\chi}^T \bmat{M}^\alpha \bvec{\chi},
\end{equation}
with $\bvec{\chi}$ being a real Majorana four-vector and
\begin{equation} \label{eq:mbasis}
  \bmat M^1 = \tau^3 \otimes \ii \tau^2, \quad \bmat M^2 = \ii \tau^2 \otimes \tau^0 \ \text{and} \ \bmat M^3 = \tau^1 \otimes \ii \tau^2
\end{equation}
suitably chosen $\SOfour$ matrices, which are to be understood as the Majorana analogue of the Pauli matrices acting on the spinor $(f_\uparrow, f_\downarrow)^T$.\cite{smv17}
Note that above representation can be seen to be equivalent to Kitaev's representation by employing the Hilbert-space constraint $D=1$.
The representation \eqref{eq:spinrep} admits a redundancy $\bvec{\chi} \to \bmat{G}^\alpha \bvec{\chi}$, where
\begin{equation} \label{eq:gbasis}
  \bmat G^1 = -\tau^0 \otimes \ii \tau^2, \quad \bmat G^2 =  -\ii \tau^2 \otimes \tau^3 \ \text{and} \ \bmat G^3 = -\ii \tau^2 \otimes \tau^1,
\end{equation}
are three $\SOfour$ matrices which commute with the $\bmat{M}^\alpha$ and form another representation of $\SUtwo$. The matrices $\bmat G^\alpha$ can be understood as an analogue of Pauli matrices for the Nambu spinor $(f_\uparrow, f_\downarrow^\dagger)^T$.
Accordingly, we can form an isospin $J^\alpha = \ii/4 \bvec{\chi}^T \bmat{G}^\alpha \bvec{\chi}$.
It can be seen that the (gauge) constraint amounts to working in the subspace of states $\ket{\psi}$ which are isospin singlets, $J^\alpha \ket{\psi} = 0$, guaranteeing that the local physical Hilbert space is indeed two-dimensional.

Kitaev's spin representation is finally obtained by considering the difference
\begin{equation} \label{eq:kitaev_rep}
  S^\alpha_\mathrm{K} \equiv \ii \chi^0 \chi^\alpha = S^\alpha - J^\alpha.
\end{equation}
When using this spin representation, it is clear that in order to realize symmetry operations acting on $S^\alpha_\mathrm{K}$, the transformation needs to act both on the physical spin sector and the isospin (gauge) sector in the same manner -- this is precisely the projective realization of symmetry operations characteristic for quantum ordered states.\cite{wen02}
For the Kitaev model, we find that an identical operation needs to act on the spin and isospin, known as ``spin-gauge locking''.\cite{ykv12}
In this case and with above choice of matrices, a joint spin- and gauge transformation
\begin{equation} \label{eq:proj_op}
  \bvec \chi \to \bmat{R}_G \bmat{R}_M \bvec \chi
\end{equation}
treats the $\chi^0$ Majorana as a scalar, while $\chi^\alpha$ transforms as a three-dimensional vector.
Importantly, the resulting mean-field Hamiltonian with a chosen ansatz also has the property that symmetries need to be realized projectively, as described above.
In the isotropic case, we also tested the spin representation \eqref{eq:spinrep} as recently used in Ref.~\onlinecite{smv17} and find qualitative (and semi-quantitative) agreement with the results obtained by using Kitaev's spin representation.


\subsection{Mean-field theory}

We first treat a decoupled layer of the model \eqref{eq:h1} in a mean-field approximation by employing the spin representation in Eq.~\eqref{eq:kitaev_rep}.
Performing the mean-field decoupling, we obtain\cite{ykv12,smv17}
\begin{align} \label{eq:kitaev_dec}
  &\mathcal{H}^{x y z}_m = -\sum_{\alpha=x,y,z} \sum_{\langle i j \rangle_\alpha} K^\alpha S^\alpha_{m i} S^\alpha_{m j} \\
  &\to \sum_{\alpha=x,y,z} \sum_{\langle ij \rangle_\alpha} K^\alpha \left[ u^\alpha_{ij} \ii \chi_i^0 \chi_j^0 + u^0_{i j} \ii \chi_i^\alpha \chi_j^\sigma - u^0_{i j} u^\alpha_{ij} \right]
\notag
\end{align}
where the $\langle i j \rangle_\alpha$ denotes a bond of type $\alpha = x,y,z$.
The real-valued Majorana-bilinear mean fields are given by
\begin{equation}
  u^0_{ij} = \langle \ii \chi^0_i \chi^0_j \rangle \ \text{and} \ u^\alpha_{i j} = \langle \ii \chi^\alpha_i \chi^\alpha_j \rangle
\end{equation}
on $\langle i j \rangle_\alpha$-links.
We assume translational invariance, such that $u^0$ and $u^\alpha$ are parametrized by their respective values on $x,y,z$-links, and choose the convention that $i\in A$ and $j\in B$ sublattice.
In the remainder, we employ the notation $u^0(\alpha)$ and $u^\alpha(\alpha)$ to denote the values of $u^{0,\alpha}$ on $\langle i j \rangle_\alpha$-links.

The resulting Majorana-bilinear Hamiltonian can then straightforwardly be diagonalized in momentum space.
The isospin singlet constraint discussed in the previous subsection is enforced on average by the use of three Lagrange multipliers $\lambda^\alpha$, however we find that for all parameter regimes discussed here, the constraint is readily satisfied for $\lambda^\alpha = 0$.

The solutions to the mean-field equations at $T=0$ are given by
\begin{subequations} \label{eq:mft_vals}
  \begin{align}
      u^0(\alpha) &= \pm \frac{1}{N} \sum_{k \in \mathrm{BZ}/2} \cos \left(\phi(\vec k) - \vec k \cdot \vec n_\alpha \right), \\
      u^\alpha(\alpha) &= \mp 0.5.
  \end{align}
\end{subequations}
for the bonds $\alpha = x,y,z$ and $\phi(\vec k) = \arg \sum_\alpha K^\alpha \eu^{\ii \vec k \cdot \vec n_\alpha}$, where  $\vec n_\alpha$ denote the lattice vectors, using the convention $\vec n_{1/2} = (\pm 1, \sqrt{3})^T /2$ and $\vec n_3 = 0$.
The solutions to mean-field equations yield a single Majorana band with a Dirac cone as well as three flat bands corresponding to the $\chi^\alpha$ Majoranas localized on $\alpha$-bonds.
There is a $\Ztwo$ freedom in choosing the global sign of the pair $u^{0,\alpha}$ on each bond as long as the relative sign between $u^0(\alpha)$ and $u^\alpha(\alpha)$ is fixed.

The mean-field theory can be related to the exact solution of the Kitaev model by noting that the mean-field parameters $u^\alpha$ essentially correspond to the $\Ztwo$ gauge field in its ground state (i.e. flux-free) configuration.
We however stress that the flat bands do not correspond to the static excitations of the gauge field.\cite{smv17}

Considering the bilayer models, the inter-layer Heisenberg interaction $\mathcal{H}_\perp$ which constitutes a quartic interaction for the Majorana fermions can be decoupled in an analogous manner to \eqref{eq:kitaev_dec}, yielding the mean-field Hamiltonian
\begin{equation} \label{eq:perp_dec}
  \mathcal{H}_\perp = - \jp \sum_{i,\alpha} \left[ \ii w^0_i \chi^\alpha_{1i} \chi^\alpha_{2i} + \ii w^\alpha_i \chi^0_{1i} \chi^0_{2i} - w^0_i w^\alpha_i \right],
\end{equation}
where the real-valued mean fields are given by \mbox{$w^{\mu}_i = \langle \ii \chi^\mu_{1i} \chi^\mu_{2i} \rangle$} for $\mu = 0, \alpha$.
Considering the Majorana four-vectors $\bvec{\chi}$, the mean-field parameters can be written in a matrix $\bmat W$. The decoupling \eqref{eq:perp_dec} corresponds to a diagonal $\bmat W$, however, also decouplings with more general $\bmat W$ are in principle possible, cf. Ref.~\onlinecite{smv17}.


The Majorana mean-field theory (MMFT) discussed above allows us to map out the phase diagram\cite{fn:temp} at $T=0$ as a function of $\jp/K$ and anisotropy $\lambda$ for the stackings illustrated in Fig.~\ref{fig:header}.
Assuming unbroken lattice translation invariance, the problem involves six chemical potentials (trivially satisfied), and $8 + 4$ real scalar mean-field parameters ($2 \times 4$ for $u^{0,\alpha}_i(\alpha)$, where $i=0,1$ is the layer index and 4 for $w^{\mu}$).
We solve the mean-field equations by means of an iterative procedure, employing a momentum-space discretization of $24 \times 24$ points.

\begin{figure}[!tb]
\includegraphics[width=.8\columnwidth,clip]{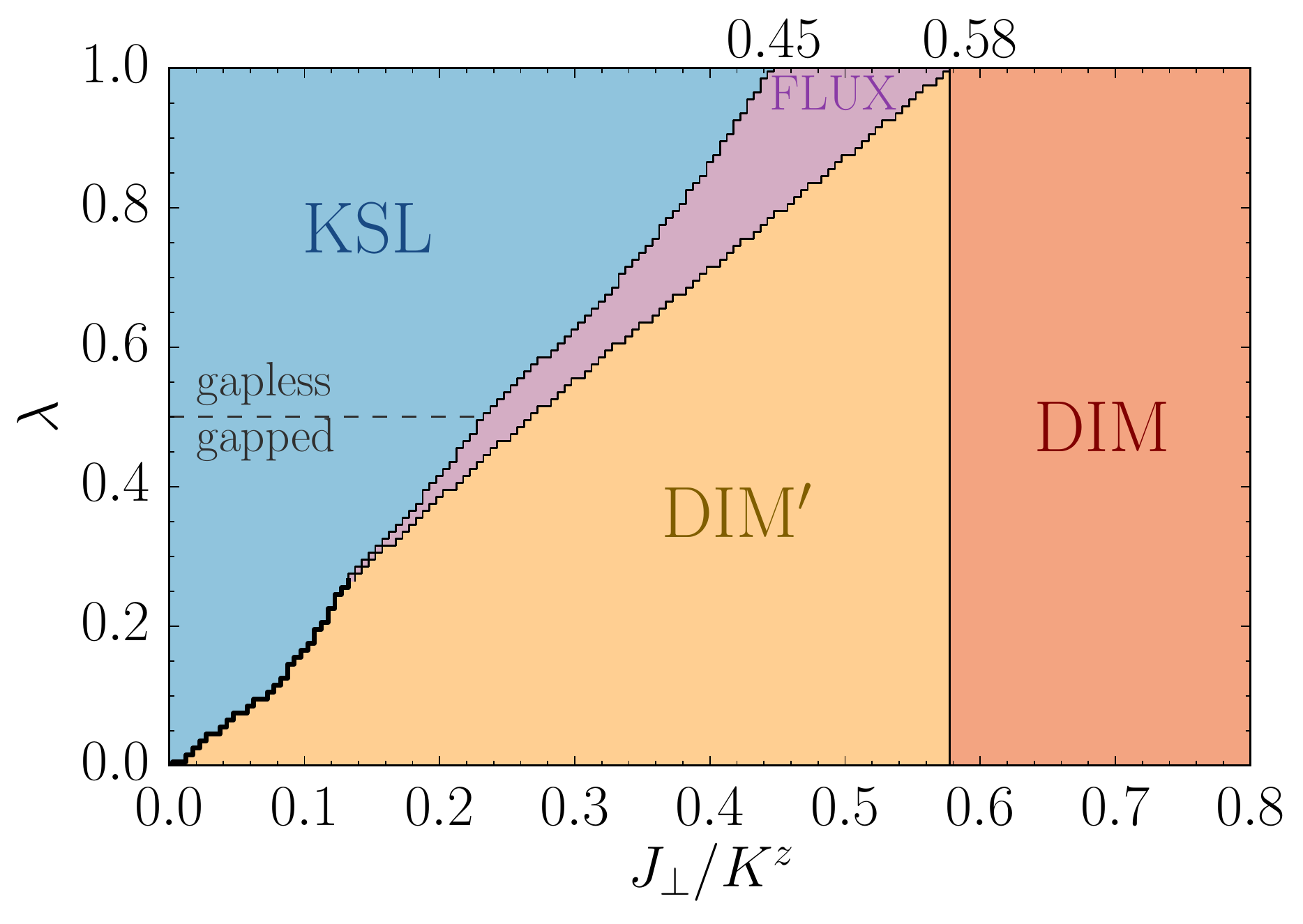}
\caption{
Mean-field phase diagram for the AA stacking. First (second) order transitions are marked with thick (thin) lines.
The \pdksli{} phase becomes gapped when $\lambda < 0.5$ (marked by a dashed line), while all other phases occurring are gapped for all parameter regimes.
 As discussed in Sec. \ref{ssec:mft_aa_stack}, we consider the \pdcross{}-phase to be an artifact of mean-field theory, and the transition at $\jp / K^z = 0.58$ can be expected to become a crossover when going beyond mean-field theory.
}
\label{fig:mft_pd_aa}
\end{figure}

\subsection{Results for the AA stacking} \label{ssec:mft_aa_stack}

\begin{figure}[!tb]
  \includegraphics[width=\columnwidth,clip]{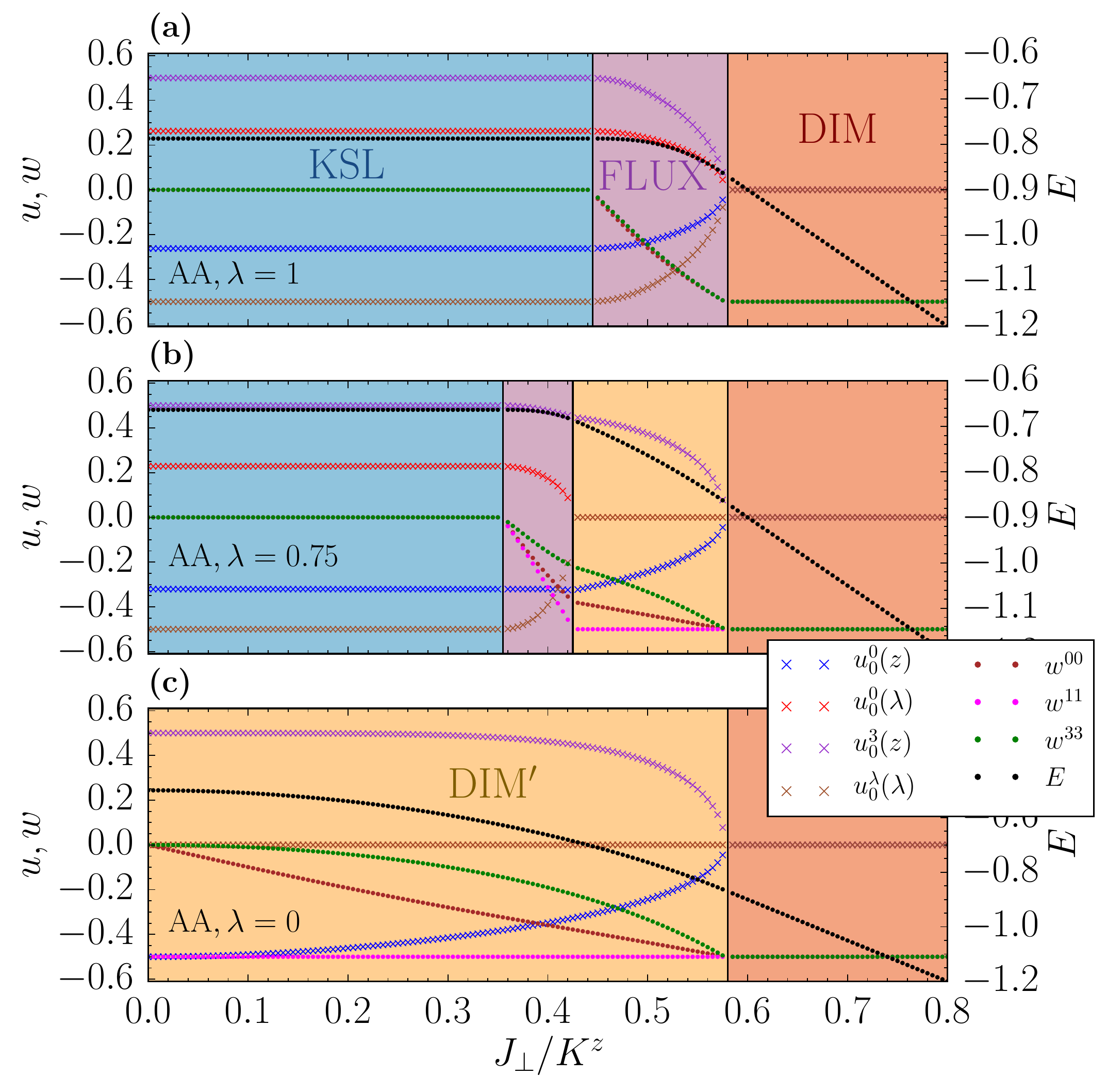}
  \caption{Mean-field parameters \cite{fn:params} obtained from the Majorana mean-field theory as a function of $\jpk$ with Kitaev couplings $K^x = K^y = \lambda K^z$. In the interest of clarity we only show MFT parameters $u^{0,\alpha}_0$ in the lower layer, since, employing a $\Ztwo$ redundancy, $u^{0,\alpha}_1 = -u^{0,\alpha}_0$. We denote the weak $x,y$-bonds and corresponding Majorana-flavors with $\lambda$.
  Also note that $w^{11} = w^{22}$ for the chosen anisotropy. (a), (b), (c) AA stacking with varying anisotropies $\lambda=0,0.75,1$. Close to the isotropic point, a phase (\pdkslii{}) with interlayer-coherence is observed. At strong anisotropies, the \pdcross{} dominates, which is separated from \pddim{} by a second-order transition which is expected to become a crossover beyond mean-field theory.
\label{fig:jpcuts}}
\end{figure}

The phase diagram as obtained from MMFT for the AA stacking is shown in Fig.~\ref{fig:mft_pd_aa}, with four mean-field phases to be discussed below.
A plot of mean-field parameters as a function of $\jpk$ for various values of $\lambda$ is shown in Fig. \ref{fig:jpcuts}.

At $\jp \ll K$ and for all $\lambda$, we find a phase for which the $u^0,u^\alpha$ mean fields are equal to the Kitaev spin-liquid parameters as shown in \eqref{eq:mft_vals}, and $w^\mu = 0$ holds.
On a mean-field level, the phase labelled \pdksli{} in Fig.~\ref{fig:mft_pd_aa} is thus identical to the decoupled limit $\jp = 0$.
Beyond mean-field theory, we expect this phase to be adiabatically connected to the $\jp = 0$ limit. Crucially, the nodes in the spectrum of the matter Majorana are protected against the perturbation $\mathcal{H}_{\perp}$, see also Sec.~\ref{sec:pert_ksl}.

Due to our parametrization of the anisotropy, decreasing $\lambda$ implies a lowering of the global energy scale for the \pdksli{}, so that the critical $\jpk$ for any transition out of the \pdksli{} phase is expected to decrease as $\lambda$ decreases.

As we increase $\jpk$ for anisotropies with $\lambda \gtrsim 0.27$, we encounter a second-order transition to a phase labelled \pdkslii{}.
In contrast to \pdksli{}, the inter-layer Heisenberg mean-fields in this phase are finite and of the form
\begin{equation}\label{eq:aniso_w}
  w^\mu = (w^{0}, w^a,w^a,w^b).
\end{equation}
The Kitaev mean-field parameters $u^{0,\alpha}\neq0$ attain numerically different values compared to the previous phase, however still preserve a structural similarity to the values in \pdksli{}, and thus can be seen to emerge continuously from the $\jp = 0$ limit.
The fact that the Kitaev mean-fields are only renormalized indicates that the quantum order (by which we refer the projective realization of symmetries, cf. Sec. \ref{subsec:majrep}) of the spin liquid is preserved.

The structure (with $w^b \neq w^a$) is a direct result of the spoiled rotational symmetry in the presence of anisotropy.
Choosing the anisotropy on a different bond, the solutions would be related by a PSG transformation which, as noted in Sec. \ref{subsec:majrep}, effectively treats the $\chi^0$ Majorana as a scalar and the $\chi^\alpha$ as a three-component vector.
In the case of $\lambda = 1$, the rotational symmetry is restored, with $w^b = w^a$.

We observe that, crucially, the $u^{0,\alpha}$ mean-field parameters in \pdkslii{} have opposite signs on the two layers, implying a breaking of the global point inversion symmetry.
Note that we can perform a $\Ztwo$ gauge transformation on one sublattice such that the $u^{0,\alpha}$ have identical signs on both layers, however this transformation leads to an alternating $w^\mu$ on the $A$- and $B$-sublattices, thus again breaking inversion symmetry.
As inversion symmetry is broken, the nodal points are no longer protected, and the presence of finite Heisenberg inter-layer mean fields $w^\mu$ implies the opening of the gap of the itinerant Majorana mode in \pdkslii{} for all $\lambda$, in contrast to \pdksli{} which is gapless for $\lambda > 0.5$.

The peculiar sign structure of the mean-field parameters implies that the itinerant $\chi^0$-Majorana fermion picks up a $\pi$-flux when going around an elementary four-spin plaquette involving two inter-layer dimers and a bond from each layer.
For a further discussion of the stability of \pdksli{} and the emergence of \pdkslii{} we refer the reader to Sec.~\ref{sec:pert_ksl}.

Upon increasing $\jpk$ further for $\lambda=1$, a second-order transition from FLUX to \pddim{} occurs: In \pddim{}, the mean-field parameters $u^0,u^a \to 0$ and $w^\mu = -\frac{1}{2} (1,1,1,1)$, and consequently all Majorana fermions transform purely by means of physical transformations, $\bvec \chi \to \bmat{R}_M \bvec \chi$, and $\bmat W \propto \bmat{1}$ is the only ansatz compatible with all symmetries. In this phase the Kitaev spin-liquid physics is completely absent, and all symmetries are preserved -- thus the mean-field ansatz transforms as a trivial representation of $\SUtwo$. Further $u^{0,\alpha} = 0$ implies that the Hamiltonian $\mathcal{H} = \mathcal{H}_\perp$ is local and portrays singlet formation between the local moments in the two layers, as expected in the limit $\jp / K \gg 1$.

At all $\lambda < 1$, an intermediate \pdcross{} phase appears, bounded by a first-order-transition into \pdksli{}, a second-order transition into \pdkslii{}, and a second-order transition into \pddim{}.
The phase \pdcross{} features vanishing $u^{0,\alpha}$ parameters on the weak $x$- and $y$-bonds, and finite values on the $z$-bonds. The inter-layer mean fields are now of the form
\begin{equation} \label{eq:w_aniso_2}
  w^\mu = \left(w^0,-\frac{1}{2},-\frac{1}{2},w^b \right),
\end{equation}
with $w^0,w^b \to -1/2$ as we approach the transition to the \pddim{} phase for $\jp \gg K^z$.
Choosing an anisotropy on a different link type results in a mean-field solution where the last three components in Eq.~\eqref{eq:w_aniso_2} are permuted accordingly.
We note that the nature of the mean-field parameters $w^\mu$ can be understood by considering that at $\lambda = 0$ (cf. Fig.~\ref{fig:jpcuts}), the $x$-and $y$-Majoranas in the Kitaev Hamiltonian constitute zero modes. Turning on a finite $\jp$, these become localized modes on the interlayer dimers with $\langle \ii \chi_1 \chi_2 \rangle = - 1/2$, as also obtained numerically in Eq.~\eqref{eq:w_aniso_2}.

In the following we argue that the \pddim{}--\pdcross{} transition, occurring at $\jpk\approx0.58$ independent of $\lambda$, is an artifact of mean-field theory. Consider first $\lambda=0$. Here we can analyze the eigenenergies and eigenstates the four-spin Hamiltonian
\begin{equation}
  \mathcal{H}_\perp = \jp (\vec S_1 \cdot \vec S_2 + \vec S_3 \cdot \vec S_4) - K^z (S_1^z S_3^z + S_2^z S_4^z).
\end{equation}
For $K^z = 0$ the ground states are trivially given by two spin singlets. As we turn on a finite $K^z$, we observe a continuous evolution of the ground state to the $K^z / \jp \gg 1$ limit without signs of an (avoided) level crossing, thus showing a crossover behavior. The finite gap implies that this behavior persists to finite $\lambda$. Hence, the second-order transition between \pdcross{} and \pddim{} observed at $\jp \simeq 0.58$ is a mean-field artifact, and both \pddim{} and \pdcross{} should be considered to represent a single dimer phase adiabatically connected to the $\jpk\to\infty$ limit.

\subsection{Results for AB stacking}
\label{ssec:mft_ab_stack}

\begin{figure}[!tb]
\includegraphics[width=.8\columnwidth,clip]{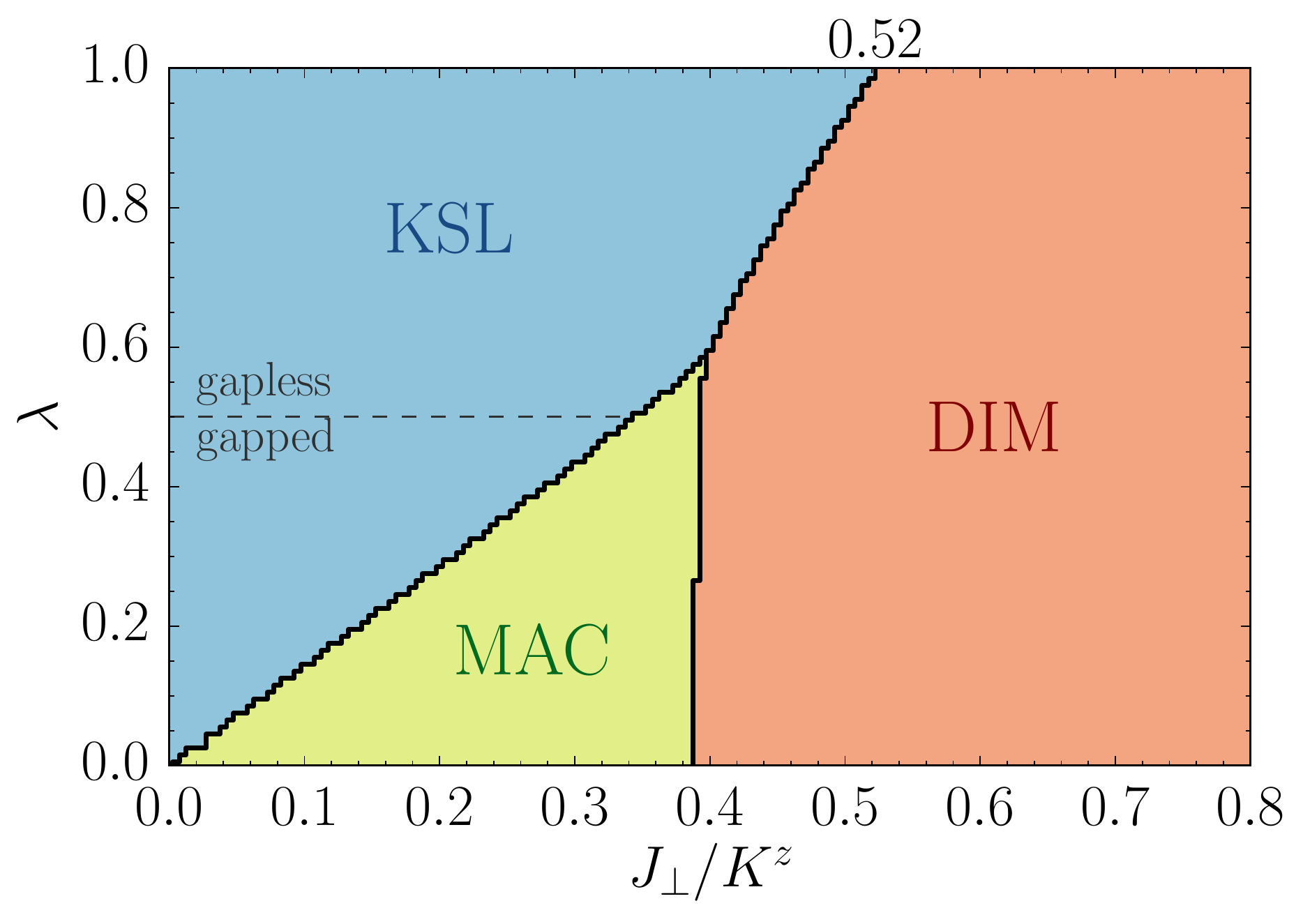}
\caption{Mean-field phase diagram for the AB stacking with an anisotropy $K^x = K^y = \lambda K^z$. First- \mbox{(second-)} order transitions are marked with thick (thin) lines. \label{fig:mft_pd_ab}}
\end{figure}

The quantitative mean-field phase diagram for the AB stacking is shown in Fig. \ref{fig:mft_pd_ab}. We discuss the three occurring phases below.

At $\lambda \gtrsim 0.58$, we find a first-order transition between the spin liquid \pdksli{} to the dimer phase \pddim{}, with the Heisenberg mean fields $\bmat W$ vanishing in \pdksli{} and taking a uniform form $w^\mu = \pm 1/2$ (as for the AA stacking), respectively.
The values of the mean fields in vicinity of the transition in each phase are identical to those at the limits $\jp = 0, K^\alpha \neq 0$, and $K^\alpha = 0, \jp \neq 0, $ respectively.
The critical $\jpk \simeq 0.52$ is thus fully determined by the energetics of the decoupled QSL and dimer phases, respectively.

\begin{figure}[!tb]
  \includegraphics[width=\columnwidth, clip]{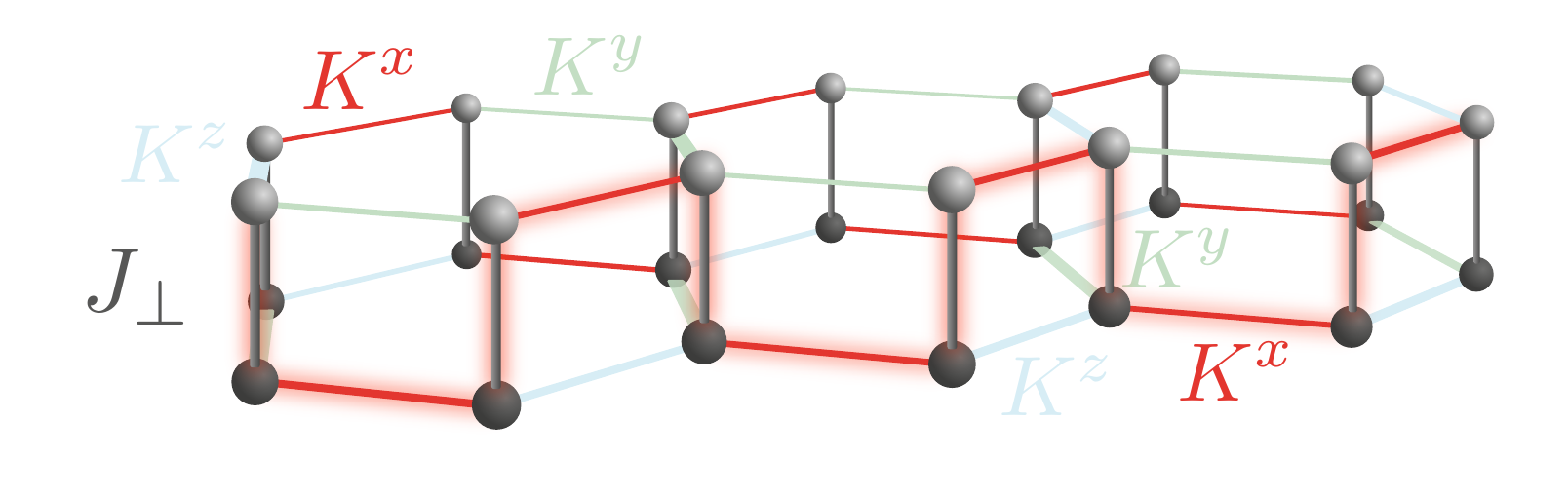}
  \caption{Illustration of an effective zigzag chain in the \pdchains{} phase for the AB-stacked model with strong anisotropy on the $x$-bonds (for visual clarity, a different anisotropy was chosen than in the main text).}
  \label{fig:chains_illu}
\end{figure}

\begin{figure}[!tb]
  \includegraphics[width=\columnwidth,clip]{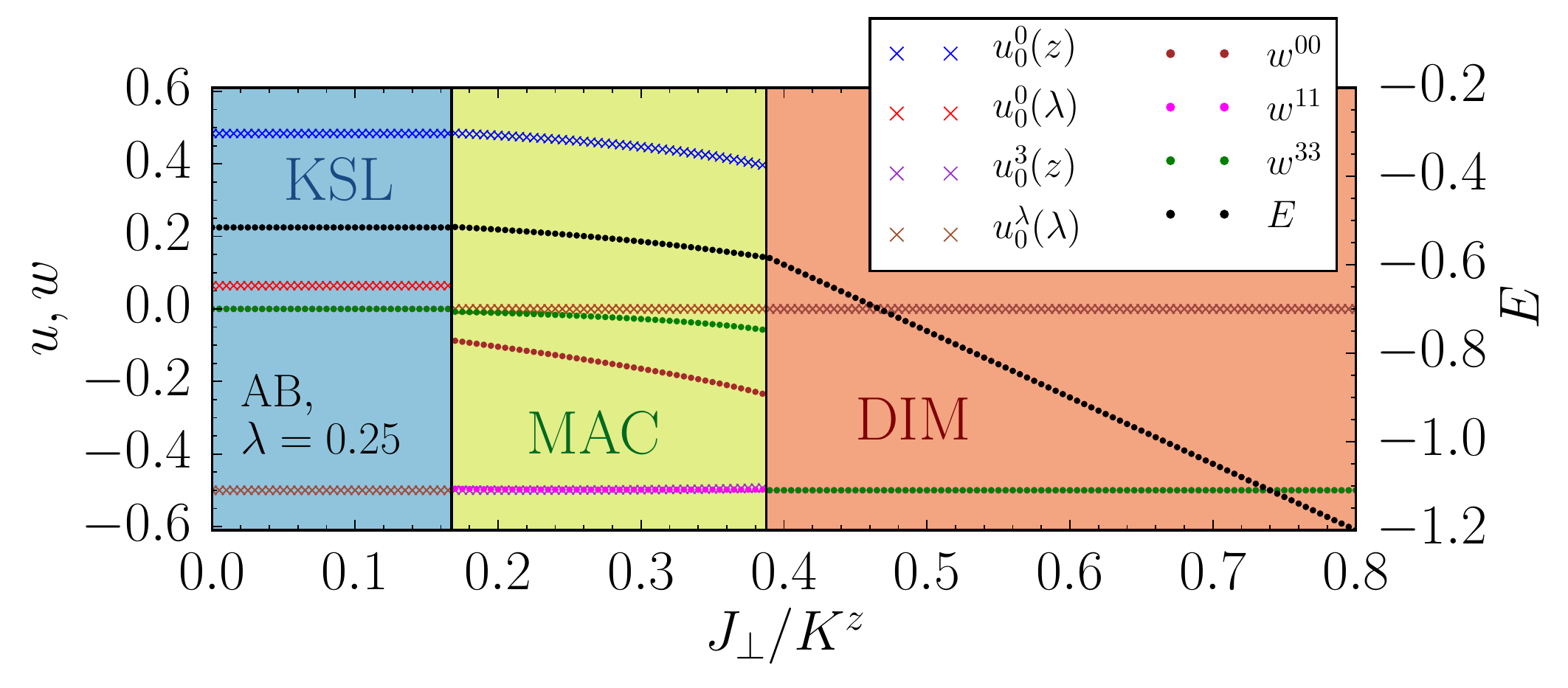}
  \caption{Mean-field parameters \cite{fn:params} for the AB stacking obtained from the Majorana mean-field theory as a function of $\jpk$ with anisotropic Kitaev couplings $K^x = K^y = \lambda K^z$ with $\lambda = 0.25$. The mean-field parameters in the \pdchains{}-phase describe decoupled chains. Since the chains are decoupled on mean-field level, the MFT parameters for  \pdchains{} in the AC-stacking are identical.
\label{fig:onejpcut}}
\end{figure}

Below $\lambda \simeq 0.58$, however, an intermediate phase, which we call \pdchains{}, emerges. A plot of the evolution of mean-field parameters as a function of $\jpk$ is shown in Fig.~\ref{fig:onejpcut}.
Remarkably, the critical $\jp$ for the transition between \pdchains{} and \pddim{} is only weakly dependent on $\lambda$ and extends down to the limiting case of $\lambda = 0$ at $\jp \simeq 0.39$. 
This is in stark contrast to the previously discussed AA stacking, for which in the anisotropic limit an infinitesimal $\jp$ suffices to enter the \pdcross{} phase (which is to be considered part of the \pddim{}-phase beyond mean-field theory).

Considering the anisotropic limit, we note that the model now effectively consists of chains formed from the strong dimers in the upper and lower layer, connected via the Heisenberg interaction, as shown in Fig.~\ref{fig:chains_illu}.

This chain can be viewed as an effective one-dimensional (1D) hopping problem for the itinerant Majorana fermions.
Decreasing the anisotropy, i.e. allowing a finite $\lambda > 0$, would result in an effective coupling of the chains.

\begin{figure}[!tb]
  \includegraphics[width=\columnwidth, clip]{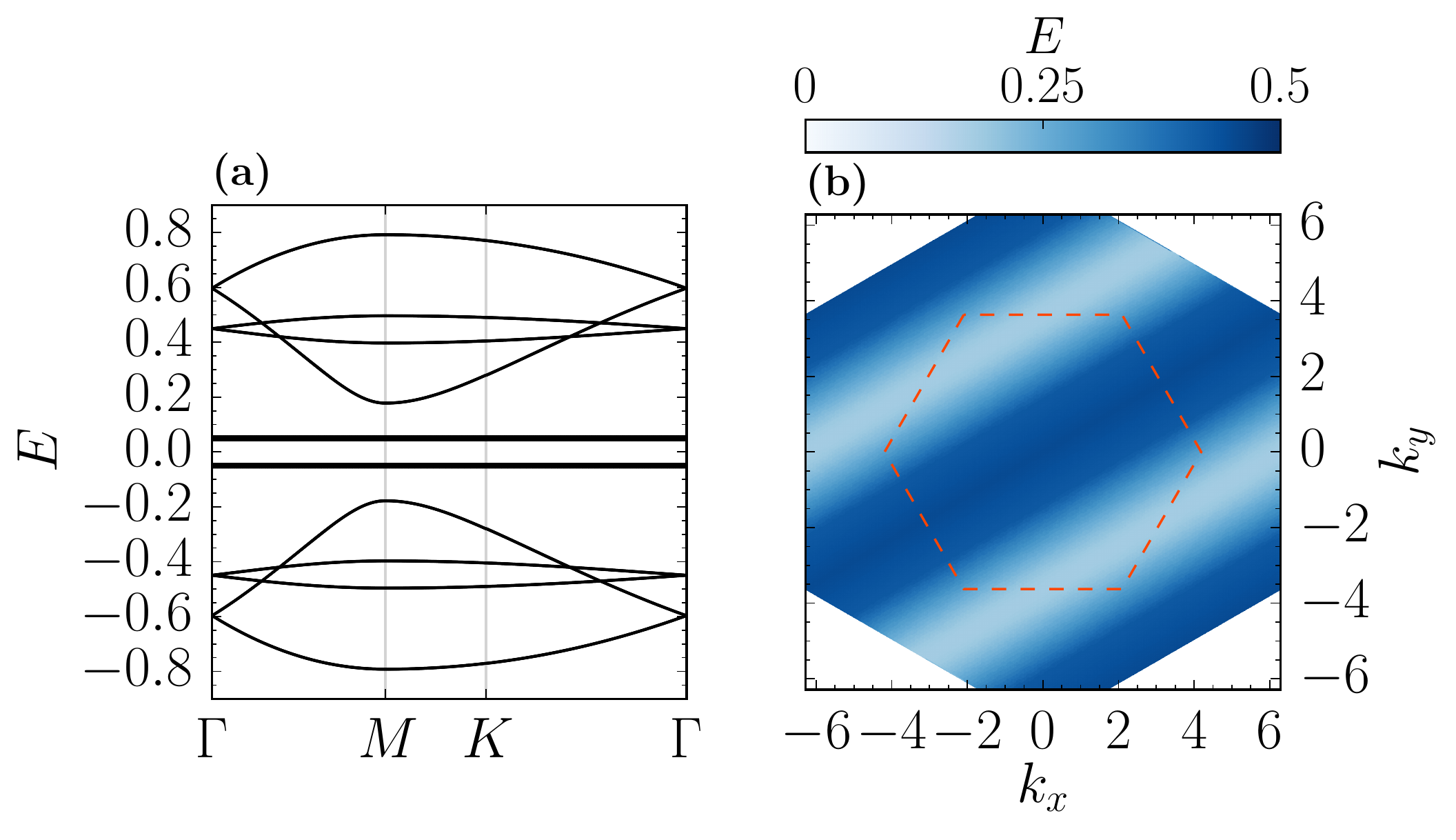}
  \caption{Band structure in the \pdchains{} phase for the AB-stacked model at $\jp = 0.3$ and $\lambda = 0.28$, showing a one-dimensional character due to hopping on chains. (a) Cut along high-symmetry lines. (b) Energy of lowest \emph{dispersing} quasiparticle band (obtained by removing flat bands resulting from localized excitations)}
  \label{fig:mft_chains_disp}
\end{figure}

In mean-field theory however, finite (but small) mean-field parameters are only induced on the $z$-links in the lower and $y$-links in the upper layer, i.e. those links which would complete the chains to ladders, but without inter-chain couplings.
In particular, we note that finite $u^0_0(z)$ and $u^0_1(y)$ are induced even in the $K^y = K^z = 0$ limit, while the finite $u^3_0(z)$ and $u^2_1(y)$ can be seen only to emerge when $\lambda > 0$, i.e. the finite amplitudes for the localized Majoranas are induced by the itinerant Majoranas on the bonds.
The Heisenberg mean fields $w^\mu$ take the form $w^\mu= (w^0, -1/2, -1/2, w^b)$, where $|w^b| < |w^0|$.

The effective one-dimensional character of this phase is also evident in the spectrum, illustrated in Fig. \ref{fig:mft_chains_disp} by a cut along high-symmetry lines and a plot of the lowest quasiparticle energy (after removing low-lying flat bands resulting from the localized Majoranas).
Notably we find that the spectrum is gapped for all parameter regimes in the \pdchains{} phase.

This phase is further discussed in Sec.~\ref{sec:novel}, where we also derive effective models by mapping the chains to effective macro-spins.


\subsection{Results for the \sac{} and \sbac{} stackings}

\begin{figure}[!tb]
\includegraphics[width=.8\columnwidth,clip]{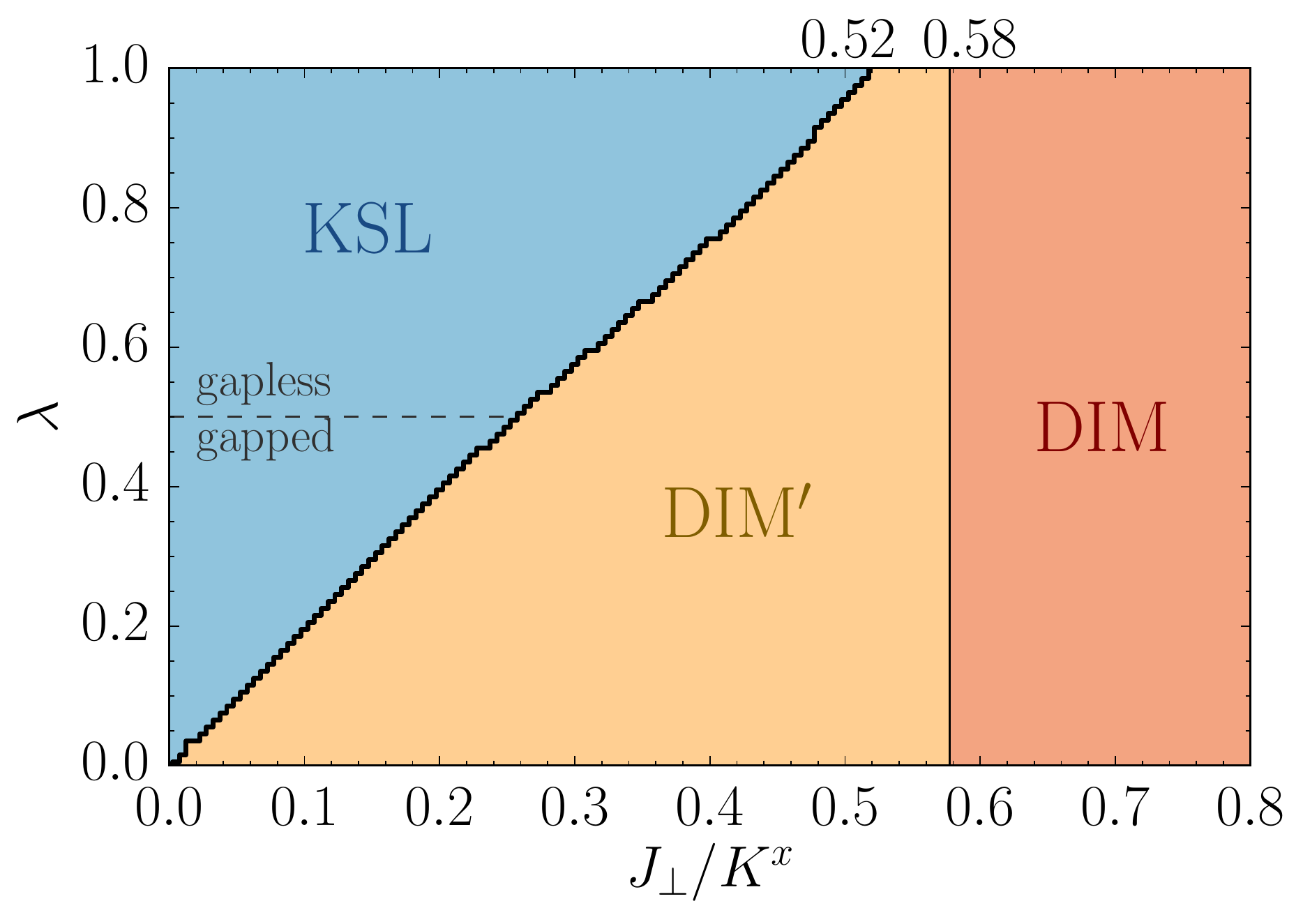}
\caption{Mean-field phase diagram for the \sac{} stacking with a symmetry-compatible anisotropy $K^y = K^z = \lambda K^x$. First- (second-) order transitions are marked with thick (thin) lines.\label{fig:mft_pd_ac}}
\end{figure}

\begin{figure}[!tb]
\includegraphics[width=.8\columnwidth,clip]{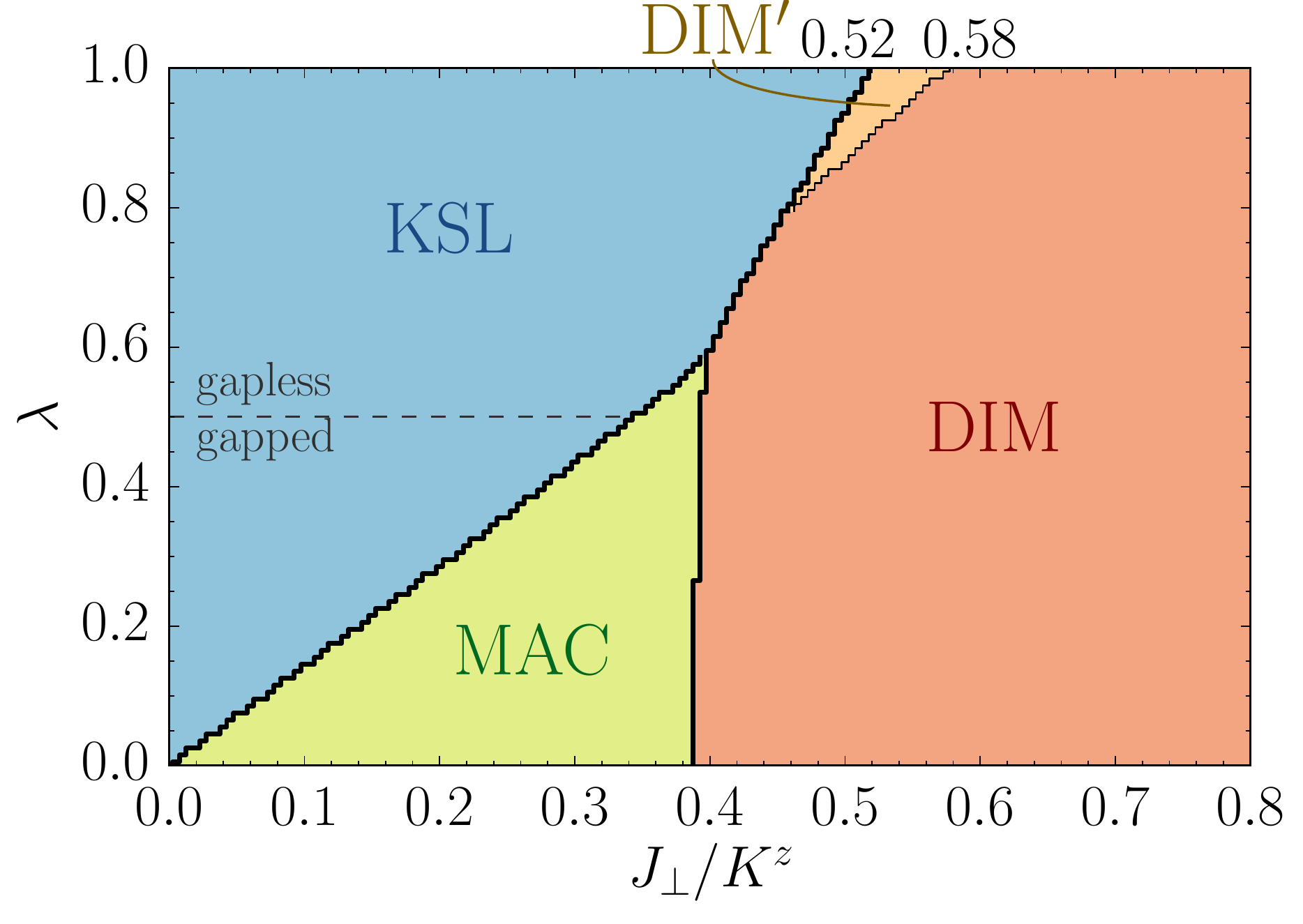}
\caption{Phase diagram for the \sbac{} stacking with anisotropy $K^x = K^y = \lambda K^z$ which spoils the reflection symmetry $\sigma$ of the AC-stacking. First- (second-) order transitions are marked with thick (thin) lines.\label{fig:mft_pd_bc}}
\end{figure}

The respective phase diagrams for the \sac{}- and \sbac{}-stacked models are shown in Figs.~\ref{fig:mft_pd_ac} and \ref{fig:mft_pd_bc}.

In the isotropic case $\lambda = 1$, these phase diagrams coincide.
There is an intermediate phase with non-finite mean-field parameters only on the $x$-bonds.
This phase is identical to phase \pdcross{} previously discussed for the AA stacking which is separated from the Kitaev spin liquid by an first-order transition at $\jpk \simeq 0.52$ and from the dimer phase by a second-order phase transition at $\jpk \simeq 0.58$.
The models become inequivalent upon decreasing $\lambda < 1$.

\subsubsection{\sac{} stacking.} When introducing the anisotropy on the $x$-bonds, compatible with the mirror symmetry $\sigma$, we observe that the first-order transition between phases \pdksli{} and \pdcross{} extends down to $\jp = 0$ as $\lambda \to 0$, as visible in Fig.~\ref{fig:mft_pd_ac}.
In addition we find that the critical $\jp$ for the second-order transition separating \pdcross{} and the dimer phase is independent of $\lambda$, for the same reasons as explained in the AA stacking.
Moreover, the mean-field parameters in \pdcross{} are identical to those in the phase \pdcross{} in the AA stacking, such that we employ the same reasoning as above to conclude that \pdcross{} is an artifact of our mean-field theory and should be associated with the dimer phase.

\subsubsection{\sbac{} stacking.}  At $\lambda < 1$ we find that the critical $\jp$ for both the the transition from \pdksli{} to intermediate phase \pdcross{} and from \pdcross{} to \pddim{} are lowered -- this is in contrast to the previous case, where the transition from phase \pdcross{} to the \pddim{} is independent of the anisotropy $\lambda$.
The fact that the \pdcross{}-\pddim{} transition depends on $\lambda$ is due to the fact that an anisotropy $\lambda < 1$ now weakens the $x$-bonds, and thus has an influence on the energetics of the \pdcross{}-phase in the \sbac{} stacking, thus also influencing the critical $\jpk$ for this (mean-field) transition.

At $\lambda \simeq 0.78$ the phase \pdcross{} eventually terminates, yielding a first-order transition between the \pdksli{} and dimer phases (cf. Fig.~\ref{fig:mft_pd_bc}).
A further intermediate phase emerges at $\lambda \simeq 0.58$, with mean-field parameters being identical to those of the \pdchains{}-phase (modulo necessary permutations relating the different stackings) obtained in the AB-stacked model (cf. above).

As for the AB stacking, the emergence of this phase can be elucidated by an effective model of chains with the links being the strong bonds, alternating between upper and lower layer.
We note that below $\lambda \simeq 0.58$, the phase diagram is fully equivalent to the phase diagram in the AB stacking.
We emphasize that in the AC stacking discussed here the couplings between the chains are only of $K^x$-type (as opposed to AB stacking, where the chains are coupled both via $K^y$ and $K^x$), however the corresponding mean-field parameters for coupling the chains vanish, such that the mean-field phenomenology for this phase is identical to the AB stacking.


\section{Series expansion}
\label{sec:series_expansion}

While the Majorana mean-field theory described above is exact in the limit $\jpk = 0$, it is not expected that results regarding the location and critical properties of phase transitions in the bilayer Kitaev model are quantitatively reliable.
However, progress can be made by considering series expansions starting from exactly known limits.

In this work we perform two series expansions. The first is about the limit of isolated $J_\perp$-dimers, i.e.~in the dimer phase where the non-degenerate ground state is adiabatically connected to the product state of singlets for $J_\perp=0$ and excitations corresponds to spin-one triplons (dressed triplets).\cite{schmidt03}
The goal of this expansion is to extract expressions for the ground-state energy and the one-triplon dispersion up to high orders in perturbation.
The second expansion is about the limit of isolated $K^z$-Ising dimers (equivalently about $K^x$- or $K^y$-Ising dimers).
Here the ground state is extensively degenerate and the purpose of the expansion is to derive an effective low-energy theory for the anisotropic limit of the Kitaev models, which results in two topologically-ordered Wen plaquette models coupled by the inter-layer Heisenberg exchange $J_\perp$.
Technically, both high-order expansions can be realized with the help of perturbative continuous unitary transformations (pCUTs) \cite{knetter00,knetter03} and we describe its generic aspects in the following.

One can always rewrite any Hamiltonian ${\cal H}$ \emph{exactly} as
\begin{equation}
\label{Eq:Hami}
{\cal H}={\cal H}_0+\sum_{j=1}^{N_\lambda}\lambda_j {\cal V}^{(j)} \quad ,
\end{equation}
where the sum runs over appropriate supersites and the $\lambda_j$ are the perturbative parameters. For the bilayer Kitaev model we use two different dimers, $J_\perp$-dimers or $K^z$-dimers, as supersites which have an equidistant spectrum bounded from below.

The unperturbed part of $\mathcal{H}$ is diagonal in the dimers $i$ of the lattice and can be written as
\begin{eqnarray}\label{h_0_q}
\mathcal{H}_0 &=& E_0+\mathcal{Q} \quad ,
\end{eqnarray}
where $E_0$ denotes a constant and $\mathcal{Q}$ is a counting operator of local excitations. This decomposition of $\mathcal{H}_0$ is always possible as long as the local spectrum of a supersite is equidistant.

Supersites interact via the perturbation \mbox{$\mathcal{V}\equiv\sum_j\lambda_j\mathcal{V}^{(j)}$}. For the bilayer Kitaev model the perturbation ${\cal V}$ couples two dimers in each of the two expansions. As a consequence of Eq.~\eqref{h_0_q}, one can rewrite Eq.~\eqref{Eq:Hami} as
\begin{equation}
\label{Eq:Hami_final}
{\cal H}={\cal H}_0+ \sum_{n=-N}^N \hat{T}_n \quad ,
\end{equation}
so that $[\mathcal{Q},\hat{T}_n]=n\hat{T}_n$. Physically, the operator \mbox{$\hat{T}_n \equiv\sum_j \lambda_j \hat{T}^{(j)}_n$} corresponds to all processes where the change of energy quanta with respect to $\mathcal{H}_0$ is exactly $n$. The maximal (finite) change in energy quanta is called $\pm N$. For the bilayer Kitaev model $N=2$ in both expansions.

In pCUTs, Hamiltonian \eqref{Eq:Hami_final} is mapped model-independently up to high orders in perturbation to an effective Hamiltonian $\mathcal{H}_\text{eff}$ with $[\mathcal{H}_{\rm eff},\mathcal{Q}]=0$. The general structure of $\mathcal{H}_{\rm eff}$ is then a weighted sum of operator products $\hat{T}_{n_1}\cdots \hat{T}_{n_k}$ in order $k$ perturbation theory. The block-diagonal $\mathcal{H}_\text{eff}$ conserves the number of quasi-particles (qp). This represents a major simplification of the quantum many-body problem, since one can treat each quasi-particle block, corresponding only to a few-body problem, separately.

The more demanding part in pCUTs is model-dependent and corresponds to a normal-ordering of $\mathcal{H}_\text{eff}$ for which the explicit processes of ${\cal H}_0$ and $\mathcal{V}$ have to be specified.
This is most efficiently done via a full graph decomposition in linked graphs using the linked-cluster theorem and an appropriate embedding scheme afterwards.
The details of the two expansions can be found in Secs.~\ref{sec:series_expansion_dimer} and \ref{sec:qpt_eff_mod_aa}.


\section{Series expansion in the dimer paramagnet}
\label{sec:series_expansion_dimer}

In this section we list our findings for the perpendicular-dimer pCUT starting from the limit \mbox{$\jpk \gg 1$}.
In the following, we set $\jp \equiv 1$ for convenience.
Our results are complemented by bond-operator theory as detailed in Appendix~\ref{sec:BOT}.

As from Sec.~\ref{sec:series_expansion}, the ground-state energy is obtained from $\langle 0
| \mathcal{H}_\text{eff} | 0 \rangle$, where $|0 \rangle = \prod_{l} | s_{l}
\rangle$ is the product state of isolated $J_\perp$-dimers. For all stackings
AA, AB, and AC we have obtained $O(9)$ expansions of type $E_0 = \sum_{l+m+n\leq
9} a_{l,m,n} K^{x,l} K^{y,m} K^{z,n}$. Table \ref{tab:series} displays the
coefficients of these series in the isotropic limit, i.e. $K^x = K^y = K^z
\equiv K$ for all stackings. As can be read off from this table, vacuum
fluctuations for AA are strongest, leading to the largest corrections to the
ground state energy. For the AB stacking, e.g., only the quadratic term is
significant.

For the one-particle excitations, i.e.~$Q{=}1$, we use that for the effective
Hamiltonian the parity, i.e.~the type $\alpha{=}x, y,z$ of the triplet is
conserved upon dispersion, and we employ translational invariance of the
honeycomb lattice with its underlying two-site basis. In turn all dispersions
$E({\bf k})_{\alpha,\mu}$, with $\mu{=}1,2$ labeling two dispersing bands,
follow from diagonalization of 2$\times$2-matrices of type $h_{\rm eff}({\bf
k})_{\alpha,\mu\nu} = \sum_{{\bf r}_{\mu}, {\bf r}_{\nu}} e^{i{\bf k} \cdot
\delta{\bf r}_{\mu\nu}} \langle \alpha{\bf r}_{\mu} \mu| \mathcal{H}_{\rm eff} |
\alpha{\bf r}_{\nu} \nu \rangle -\delta_{ \delta{\bf r}_{ \mu \nu},0} E_{0}$,
where $|\alpha{\bf r}_{\nu} \nu \rangle$ refers to a parity-$\alpha$ triplet, on
site ${\bf r}_{\nu}$, of basis element $\nu {=}1,2$. We note, that at general
locations in ${\bf k}$-space the corresponding secular equation can imply that
$E({\bf k})_{\alpha,\mu}$ is non-analytic in $K^{x,y,z}$.

\begingroup
 \begin{table}[!tb]
 \renewcommand*{\arraystretch}{1.3}
\begin{center}
\begin{tabular}{l|ccccccc}
\multicolumn{1}{c|}{$n$} & $0$ & $1$ & $2$ & $3$ & $4$ & $5$ & $6$ \\\hline
$E_{0,\text{AA}}$ & $-\frac{3}{2}$ & $\phantom{+}0$ & $-\frac{3}{8}$ & $0$ & $\phantom{+}\frac{15}{128}$ & $\phantom{+}0$ & $-\frac{21}{256}$ \\
$E_{0,\text{AB}}$ & $-\frac{3}{2}$ & $\phantom{+}0$ & $-\frac{3}{16}$ & $0$ & $\phantom{+}\frac{1}{256}$ & $\phantom{+}0$ & $\phantom{+}\frac{1807}{1179648}$ \\
$E_{0,\text{AC}}$ & $-\frac{3}{2}$ & $\phantom{+}0$ & $-\frac{1}{4}$ & $0$ & $\phantom{+}\frac{11}{384}$ & $\phantom{+}0$ & $-\frac{10769}{1769472}$ \\
$\Delta_{\text{AA}}$ & $1$ & $-\frac{1}{2}$ & $\phantom{+}\frac{3}{8}$ & $\frac{1}{16}$ & $-\frac{27}{128}$ & $-\frac{9}{256}$ & $\phantom{+}\frac{3}{16}$ \\
$\Delta_{\text{AB}}$ & $1$ & $-\frac{1}{2}$ & $-\frac{1}{4}$ & $\frac{31}{128}$ & $\phantom{+}\frac{91}{3072}$ & $-\frac{7249}{73728}$ & $\phantom{+}\frac{8681}{589824}$ \\
$\Delta_{\text{AC}}^\text{x}$ & $1$ & $-\frac{1}{2}$ & $-\frac{1}{4}$ & $\frac{11}{32}$ & $\phantom{+}\frac{29}{192}$ & $-\frac{4025}{9216}$ & $\phantom{+}\frac{42251}{884736}$ \\
$\Delta_{\text{AC}}^\text{y/z}$ & $1$ & $-\frac{1}{2}$ & $\phantom{+}\frac{1}{16}$ & $\frac{13}{128}$ & $-\frac{5}{96}$ & $\phantom{+}\frac{1505}{73728}$ & $-\frac{11359}{294912}$
\\\hline
\multicolumn{8}{c}{}
\\
\multicolumn{1}{c|}{$n$}  & \multicolumn{2}{c}{$7$} &  \multicolumn{3}{c}{$8$} & \multicolumn{2}{c}{$9$} \\\hline
$E_{0,\text{AA}}$ & \multicolumn{2}{c}{$\phantom{+}0$} & \multicolumn{3}{c}{$\phantom{+}\frac{4941}{65536}$} & \multicolumn{2}{c}{$\phantom{+}0$} \\
$E_{0,\text{AB}}$ & \multicolumn{2}{c}{$\phantom{+}0$} & \multicolumn{3}{c}{$-\frac{1217957}{2264924160}$} & \multicolumn{2}{c}{$\phantom{+}0$} \\
$E_{0,\text{AC}}$ & \multicolumn{2}{c}{$\phantom{+}0$} & \multicolumn{3}{c}{$\phantom{+}\frac{13542397}{10192158720}$} & \multicolumn{2}{c}{$\phantom{+}0$} \\
$\Delta_{\text{AA}}$ & \multicolumn{2}{c}{$\phantom{+}\frac{281}{8192}$} & \multicolumn{3}{c}{$-\frac{13491}{65536}$} & \multicolumn{2}{c}{$-\frac{5041}{131072}$} \\
$\Delta_{\text{AB}}$ & \multicolumn{2}{c}{$\phantom{+}\frac{801589}{14155776}$} & \multicolumn{3}{c}{$-\frac{855668113}{20384317440}$} & \multicolumn{2}{c}{$-\frac{52654093663}{2446118092800}$}\\
$\Delta_{\text{AC}}^\text{x}$ & \multicolumn{2}{c}{$\phantom{+}\frac{12128615}{21233664}$} & \multicolumn{3}{c}{$-\frac{992526557}{2038431744}$} & \multicolumn{2}{c}{$-\frac{667089160007}{1223059046400}$}\\
$\Delta_{\text{AC}}^\text{y/z}$ & \multicolumn{2}{c}{$\phantom{+}\frac{3537217}{84934656}$} & \multicolumn{3}{c}{$-\frac{1302565679}{40768634880}$} & \multicolumn{2}{c}{$\phantom{+}\frac{30446086361}{815372697600}$}\\\hline
\end{tabular}
\end{center}
\caption{Expansion coefficients $c_n$ for ground-state energy $E_0$ and
energy gap $\Delta$ at BZ center $\Gamma$ in isotropic case. Expansions are
of type $\sum_n c_n K^n$.}  \label{tab:series} \end{table} \endgroup


\subsection{AA stacking}

\begin{figure}[!tb]
\includegraphics{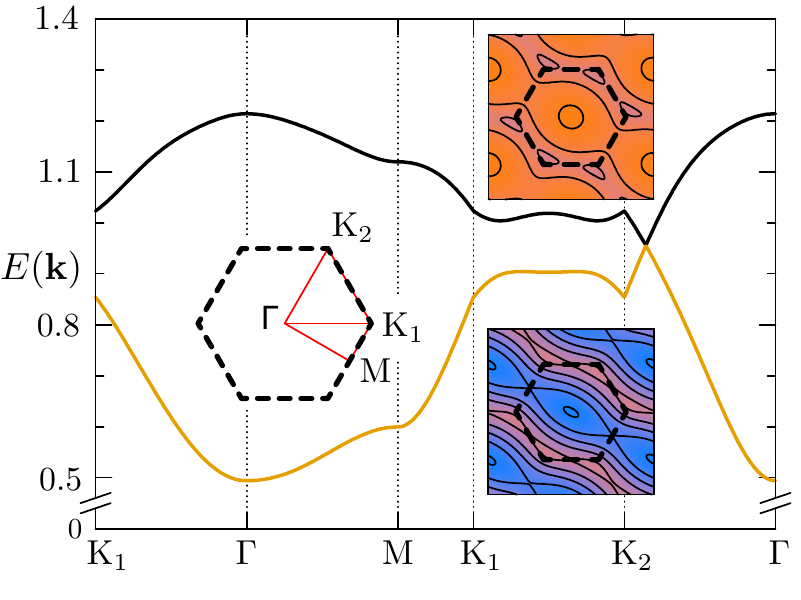}
\caption{(Color online) Dispersion $E({\bf k})_{\alpha,\mu}$ for AB stacking,
$\alpha{=}x$, and $\mu{=}1,2$ (oker,black) along high-symmetry directions in
BZ. Insets: dispersion contours for $\mu{=}1,2$ (blue,orange)}
\label{fig:dispersion}
\end{figure}

First, up to $O(9)$ and consistent with the conservation laws discussed, as well
as the bond-operator theory detailed in Appendix \ref{sec:BOT}, we find that all triplets
remain dispersionless for AA stacking.

\subsection{Isotropic AB and AC stackings} \label{sec:series_iso}

Fig. \ref{fig:dispersion} displays the dispersion of both $x$-triplons on a path
along high-symmetry directions in the BZ for $K=0.9$ at $O(9)$. As is evident, even
at this rather large inter-dimer coupling, the dispersion is strongly anisotropic,
with dominant triplet hopping along the $x$-connected zigzag-chains. While at
intermediate order of the expansion, we find exceptions, there is a stable trend for
the gap, i.e. the minimum of the dispersion to be located at the BZ center, i.e. at
${\bf k}=\Gamma$ for both AB- and AC-stacking.  For AB-stacking the $x,y,z$-triplons
are degenerate up to rotational symmetry. For AC-stacking the $x$-triplon has an energy
slightly lower than that of the $y$- and $z$-triplons and marks the gap at $\Gamma$.

\begin{figure}[!tb] \includegraphics{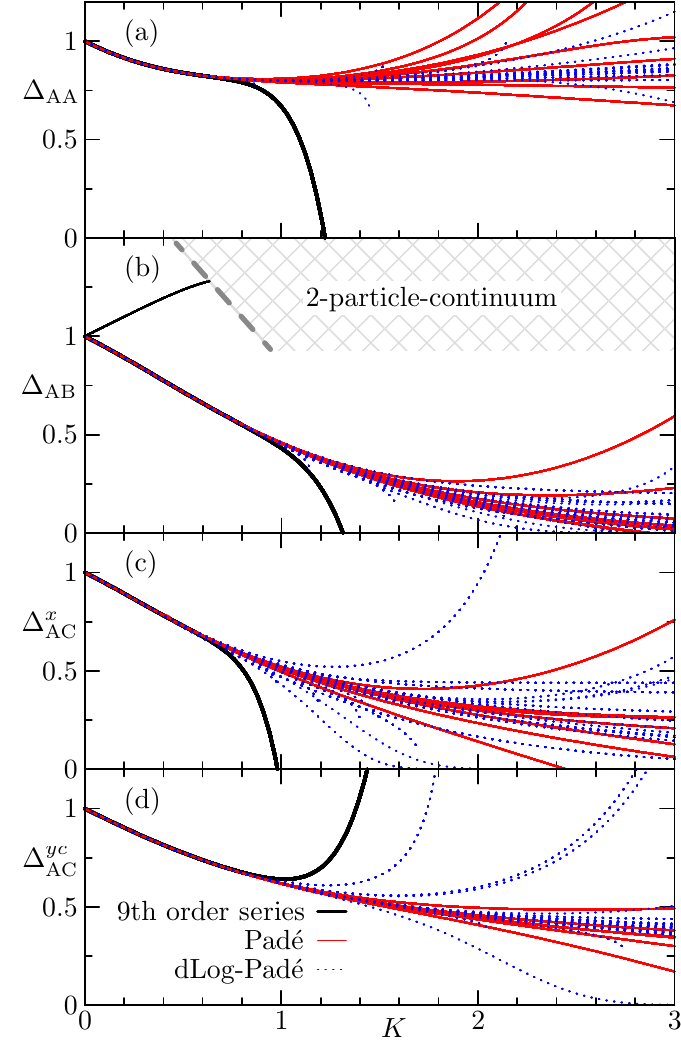} \caption{(Color online) (a-c)
$O(9)$ gap for all stackings and all distinct triplet types versus $K$ (in units of $\jp \equiv 1$) for
isotropic case: Bare gap-series (bold solid black), Pad\'e (solid red) and
dLog-Pad\'e (dashed blue). (b) Bare upper $\mu{=}2$-triplet band-gap (thin solid
black) and non-interacting two-particle continuum (gray
hatched).
}
\label{fig:pade} \end{figure}

At $\Gamma$ the secular equations for the dispersions are complete squares,
allowing to express the series for the gap $\Delta = \sum_n c_n K^n$ without
additional expansions of square roots. In Table \ref{tab:series}, the
coefficients $c_n$ are listed up to $O(9)$. In Fig. \ref{fig:pade} the gap is
analyzed in three ways:

First, the bare series is shown. In addition to that in
Fig. \ref{fig:pade}(b) the minimum of the upper triplet branch is also
depicted. As is evident from the boundary of the bare two-triplet continuum of
the lower triplet branch, also shown in this panel, the upper triplet
excitations are likely to decay into multi-particle continua and will therefore
be discarded from further discussion. All bare series depicted turn critical at
$K\sim 1$.

To assess this, we have generated order-$[m,n]$ dLog-Pad\'e
approximants to the bare gap-series for a reasonable set of $[m,n]\in [1\dots
5,1\dots 8]$. As is clear from the behavior of the majority of these
approximants in Fig. \ref{fig:pade}(a)-(d), and in stark contrast to all bare
gap-series, for {\it none} of the stackings a gap closure in the dimer phase
seems likely in the range of parameters $J_\perp/K \sim 0.5$ relevant to the
MMFT at $\lambda=1$. For the AC-stacking, higher order series would be of interest,
to further corroborate this.

Finally, Fig. \ref{fig:pade} also displays plain
$[m,n]$-Pad\'e approximants. Evidently they are very similar to the dLog-Pad\'e
approximants. This appears to be consistent with a (weak) first-order transition,
or the condensation of multi-particle modes yielding a second-order phase transition, 
as expected for a topological phase transition.

\subsection{Anisotropic AB and \sbac{} stackings} \label{sec:series_aniso}

Now we turn to the triplet gap for anisotropic coupling.
For both AB- and \sbac{}-stacking and at $\lambda=0$, we face decoupled 1D $K^z$-$J_\perp{\equiv} 1$ zigzag chains.  These exhibit an exact gap-closure $\Delta = 1-K^z/2$ at an intrachain wave vector $k_{\parallel,c}=0$, consistent with the formation of a symmetry broken macro-spin state per chain (cf. Sec. \ref{sec:novel}), showing no dispersion along straight lines connecting the $\Gamma$,$M$-points.

First, and as a direct check of our pCUT evaluation of $E({\bf k})_{\alpha,\mu}$, which in practice is of $O(9)$ in $K^{x,y,z}$, we find, that this is the case indeed.

\begin{figure}[!tb]
\includegraphics[width=1\columnwidth,clip=true]{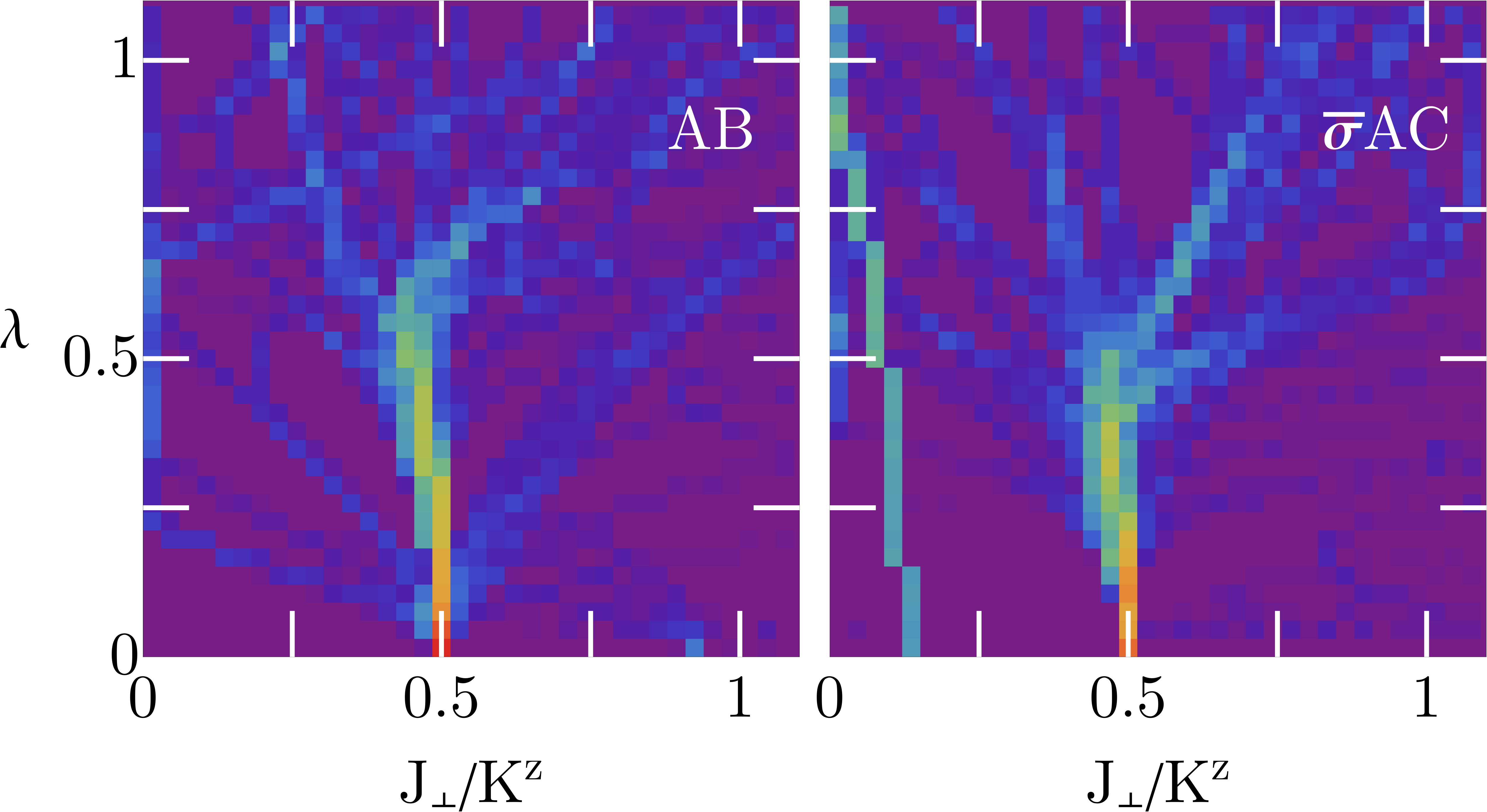}
\caption{(Color online)
Density of dLog-Pad\'e approximant poles vs. $\lambda{,}J_\perp/K^z$ for AB-
(left) and \sbac{}-stacking (right). Extracted from $K^z$-series for fixed $\lambda=K^{x,y}/K^z$.
Bin size $\delta\lambda, \delta J_\perp/K^Z \sim .03$ with color coding from single poles (dark blue) to $O(100)$ poles (green to red) per bin.}
\label{fig:pade_poles}
\label{padecontour}
\end{figure}

Second, this exact critical behavior at $\lambda=0$ can be extended to finite
$\lambda$ using dLog-Pad\'e approximants. This allows for direct comparison with the
MMFT phase diagrams from Figs. \ref{fig:mft_pd_ab} and \ref{fig:mft_pd_bc}. To
perform this analysis, and as shown in Fig. \ref{padecontour}, we scan the
($\lambda$, $J_\perp/K^Z$)-plane using $[m,n]$-dLog-Pad\'e approximants for a
reasonable set of $[m,n]\in [0\dots 8,0\dots 8]$ to a sufficiently large number of
series of a {\it single} parameter $K^z$, generated from the gap-series depending on
all of $K^{x,y,z}$, such that $K^{x,y}$ varies with $K^z$ along lines of slope
$\lambda$, with $K^x=K^y=\lambda K^z$. All pole locations of the dLog-Pad\'e
approximants are recorded in a 2D histogram, the contours of which are shown in
Fig. \ref{padecontour}.  As is very obvious from this plot, and apart from a few
spurious poles, the continuous gap-closure of the fully decoupled limit can be traced
up to $\lambda\sim 0.5$ along an essentially vertical line. This strongly
corroborated the straight line for the \pdchains{}-\pddim{} transition found in
Figs. \ref{fig:mft_pd_ab} and \ref{fig:mft_pd_bc}, although with a shift of the
transition line to $J_\perp/K^z = 0.5$. 

We note, that in contrast to the
isotropic case, only $z$-triplons are the low-energy modes, accounting for the gap for
$\lambda < 1$ in AB-stacking, as well as for $\lambda \ll 1$ in the \sbac{}-stacking. For
the latter, and to revert back to the $x$-triplon featuring the gap in the isotropic
case, the low-energy modes have to switch roles between $z$- and $x$-triplons for some
$\lambda^\prime\in[0,1]$. While the precise location of this point requires higher
orders of the pCUT, we speculate that $\lambda^\prime\approx 0.5$.

For the critical wave vector ${\bf k}_c$ of the gap closure, the series results in
two distinct scenarios. For \sbac{}, the linkage of the 1D $K^z$-$J_\perp{\equiv}
1$ zigzag chains by pairs of parallel $K^x$-bonds prevents dispersion of $y$- and
$z$-triplets transverse to the zigzag chains, identical to the lack of dispersion in
any direction for the AA stacking. Therefore, the gap-closure for the \pddim{}-\pdchains{}
transition for the \sbac{} stacking does {\it not} select a specific ${\bf k}$ point, but
continues to occur along straight lines in momentum space connecting the
$\Gamma$,M-points. This is consistent with a transition into a state with intrachain
antiferromagnetic order, but interchain degeneracy.
In contrast to this, for the AB stacking, and already at 2nd order, i.e. $O(K^x K^y)$, the series allows for triplet dispersion transverse to the zigzag chains. We find that ${\bf k}_c$=0 is selected for $\lambda\neq 0$.
This implies a non-degenerate Macro-phase for the AB-case and indicates that the nature of the \pdchains{} phase depends sensitively on the stacking.

\subsection{Stability of the \pddim{} phase} \label{sec:dimer_stab}

The fact that the triplons are strictly localized in the AA stacking suggests that the dimer phase is more stable against the effect of finite $K$, compared with the AB or \sac{}/\sbac{} stackings.
Consequently, we hence expect the critical $\jpk$ for the breakdown of the topological ordered spin-liquid phase to be smaller than in stackings with dispersing triplons, as also illustrated in the phase diagrams in Fig.~\ref{fig:header}.

%
\section{Effective plaquette models and quantum phase transition in the anisotropic AA and \sac{} stackings}
\label{sec:qpt_eff_mod_aa}

In this section we focus on the anisotropic limit of the bilayer Kitaev model with
AA stacking and we ask the question how the Abelian phases of the Kitaev model
break down when the interlayer coupling $\jp$ is turned on. To this end we derive an effective model about the dimerized limit $K^x,K^y,\jp \ll K^z$ of the
bilayer Kitaev model using the pCUT method along the lines of
Refs.~\onlinecite{schmidt08,vidal08}, as also outlined in Sec.~\ref{sec:series_expansion}.
We show that the exact local conserved quantities allow an exact duality mapping of the most relevant low-energy sector of
the effective model for the AA stacking.
This enables us to predict a second-order quantum phase transition in the (2+1)D Ising universality class between the Abelian topological phase and the trivial quantum paramagnet upon increasing $\jpk$.
%
\begin{figure}[t]
\begin{center}
\includegraphics[width=\columnwidth]{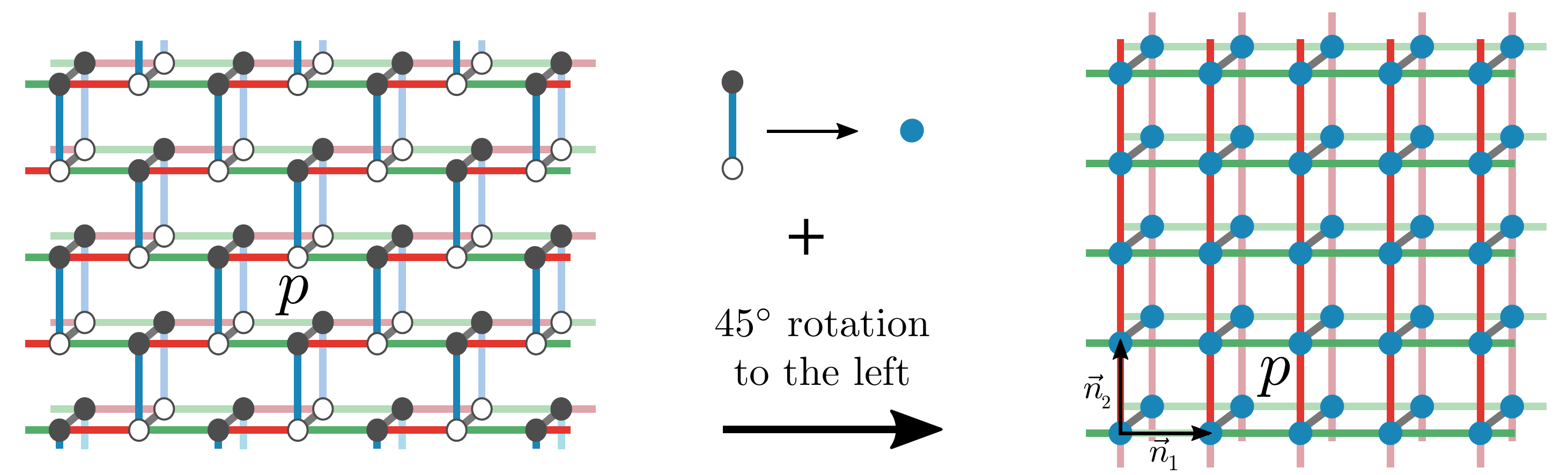}
\caption{The left picture shows the original bilayer brick wall lattice with plaquettes $p$ in each layer $m=1,2$. The spin-1/2 degrees of freedom of the bilayer Kitaev model reside on the black and white circles. The mapping replaces the two spins 1/2 degrees of each $K^z$-dimer into a hard-core boson and a pseudo-spin 1/2. These degrees of freedom reside on the blue circles of the bilayer square lattice depicted on the right side.}
\label{fig:Bilayer-Kz-mapping}
\end{center}
\end{figure}
%

\subsection{Mapping}
In the limiting case \mbox{$K^x$=$K^y$=$\jp$=0} the model is a collection of isolated $K^z$-dimers.
Each dimer has four possible configurations: two low-energy states
\mbox{$\{|\!\downarrow\downarrow\rangle,|\!\uparrow\uparrow\rangle\}$} with energy
$-K^z/4$ and two high-energy states
\mbox{$\{|\!\downarrow\uparrow\rangle,|\!\uparrow\downarrow\rangle\}$} with energy
$K^z/4$.
One can then interpret the change from a ferromagnetic to an
antiferromagnetic dimer configuration as the creation of a particle, with an
energy cost that we set equal to $1$ by choosing $K^z=2$. These particles are hardcore
bosons hopping on the sites of an effective bilayer square lattice, together with an
effective spin-$1/2$ indicating which kind of (anti)-ferro dimer configuration
is realized. We choose the following mapping \cite{schmidt08,vidal08}
%
%
\begin{equation}
  \label{eq:mappings}
  |\!\uparrow\uparrow\rangle=|\!\Uparrow\! 0\rangle,
  |\!\downarrow\downarrow\rangle=|\!\Downarrow\! 0\rangle,\,
  |\!\uparrow\downarrow\rangle=|\!\Uparrow\! 1\rangle,
  |\!\downarrow\uparrow\rangle=|\!\Downarrow\! 1\rangle,
\end{equation}
%
%
where the left (right) spin is the one of the black (white) site of the dimer,
and double arrows represent the state of the effective spin.
Let us denote by $\crb_{m\bi}$ ($\anb_{m\bi}$) the creation (annihilation) operator
of a hardcore boson at the site $\bi$ of the layer $m=1,2$ of the effective
bilayer square lattice, and $\tau_{m\bi}^\alpha$ the Pauli matrices of the effective spin at the same
site in the same layer. With these notations, the number of hardcore bosons in the system is
\mbox{$\mathcal{Q}=\sum_{m,\bi}\crb_{m\bi}\anb_{m\bi}$} and the Hamiltonian (\ref{eq:h1}) can be rewritten
as
%
%
\begin{equation}  \label{eq:ham_v2}
  H = -\frac{N}{2}+\mathcal{Q}+\sum_{\kappa\in\{{\rm K},\perp\}}\left( T^{(\kappa )}_0+T^{(\kappa )}_{+2} + T^{(\kappa )}_{-2}\right),
 \end{equation}
%
%
where $N$ is the number of $K^z$-dimers. The couplings between $K^z$-dimers in the same Kitaev layer are
 then given by
%
%
\begin{eqnarray}
 T^{({\rm K})}_0&=&-\sum_{m,\bi}\left(\frac{K^x}{4}\, t_{m,\bi}^{m,\bi+\bn_1}+\frac{K^y}{4}\, t_{m,\bi}^{m,\bi+\bn_2}
    +\mathrm{h.c.}\right),\nonumber\\
  T^{({\rm K})}_{+2}&=&-\sum_{m,\bi} \left(\frac{K^x}{4}\, v_{m,\bi}^{m,\bi+\bn_1}+\frac{K^y}{4}\, v_{m,\bi}^{m,\bi+\bn_2}\right)\nonumber\\
  &=&\left( T^{({\rm K})}_{-2}\right)^\dagger\, ,
\end{eqnarray}
%
%
with hopping and pair creation operators $t$ and $v$
%
%
\begin{eqnarray}
  \label{eq:t}
  t_{m,\bi}^{m,\bi+\bn_1}&=&\crb_{m,\bi+\bn_1}\anb_{m,\bi}\, \tau^x_{m,\bi+\bn_1}\\
  t_{m,\bi}^{m,\bi+\bn_2}&=&-\ii\, \crb_{m,\bi+\bn_2}\anb_{m,\bi}\,
  \tau^y_{m,\bi+\bn_2}\tau^z_{m,\bi}\\
  \label{eq:v}
  v_{m,\bi}^{m,\bi+\bn_1}&=&\crb_{m,\bi+\bn_1}\crb_{m,\bi}\, \tau^x_{m,\bi+\bn_1}\\
  v_{m,\bi}^{m,\bi+\bn_2}&=&\ii\, \crb_{m,\bi+\bn_2}\crb_{m,\bi}\,
  \tau^y_{m,\bi+\bn_2}\tau^z_{m,\bi},
\end{eqnarray}
%
%
and the vectors $\bn_1$ and $\bn_2$ as shown in Fig.~\ref{fig:Bilayer-Kz-mapping}.
The interaction between the two layers due to $\jp$ translates into
%
\begin{eqnarray}
 T^{(\perp)}_0&=&\frac{J_\perp}{4}\sum_{\bi}\Big[ \left( \crb_{1\bi}\anb_{2\bi} +\crb_{2\bi}\anb_{1\bi}  \right) \left( 1+\vec{\tau}_{1\bi}\cdot\vec{\tau}_{2\bi}\right) {\phantom{\Big]}}\\
          &+& {\phantom{\Big[}} 2\tau^z_{1\bi}\tau^z_{2\bi}\left[ 2\hat{n}_{1\bi}\hat{n}_{2\bi}-\left(\hat{n}_{1\bi}+\hat{n}_{2\bi}\right)\right] +2\tau^z_{1\bi}\tau^z_{2\bi}\Big]\nonumber\\
  T^{(\perp)}_{+2}&=& \frac{J_\perp}{4} \sum_{\bi} \crb_{1\bi}\crb_{2\bi}\left( 1+ \tau^x_{1\bi}\tau^x_{2\bi}+ \tau^y_{1\bi}\tau^y_{2\bi}-+ \tau^z_{1\bi}\tau^z_{2\bi}\right)\nonumber\\
         &=&\left( T^{(\perp)}_{-2}\right)^\dagger \, .
\end{eqnarray}
%
Note that the mapping from the original bilayer Kitaev model \eqref{eq:h1} to the Hamiltonian \eqref{eq:ham_v2} is exact.\cite{schmidt08,vidal08}

\subsection{Effective spin  model}

Next we apply the pCUT method\cite{knetter00,knetter03} to Eq. \eqref{eq:ham_v2}. The main idea is to transform \eqref{eq:ham_v2} which does not conserve the number of hardcore bosons into an effective Hamiltonian $H_\mathrm{eff}$ which satifies $[H_\mathrm{eff},Q]=0$. This effective Hamiltonian is a sum of $q$-quasi-particle (QP) operators with $q\in \mathbb{N}$.\cite{knetter03} Here we are only interested in the 0QP sector $q=0$ where the effective model reduces to a pure spin model in terms of the pseudo-spin degrees of freedom $\vec{\tau}$ shown as blue circles in Fig.~\ref{fig:Bilayer-Kz-mapping}. Up to order four in $K^x$, $K^y$, and $J_\perp$ we find
%
\begin{equation}\label{eq:H_eff}
\mathcal{H}^{0\textrm{QP}}_{\textrm{eff}}=\sum_{m\in\{1,2\}}\mathcal{H}^{0\textrm{QP}}_{m,\textrm{Kitaev,eff}}+\mathcal{H}^{0\textrm{QP}}_{\perp,\textrm{eff}}~,
\end{equation}
%
where the first term represents the well-known Wen-plaquette model\cite{wen03} in each Kitaev layer $m$
%
\begin{equation}\label{eq:H_eff_Kitaev}
\mathcal{H}^{0\textrm{QP}}_{m,\mathrm{Kitaev,eff}}=E_0-C_p\sum_p \hat{W}_{m,p}~,
\end{equation}
%
with the constant contribution
\begin{equation}
  32 E_0/N=-16-(\left(K^x\right)^2+\left(K^y\right)^2)-\left(\left(K^x\right)^4+\left(K^y\right)^4\right)/64,
\end{equation}
 and \mbox{$C_p=\left(K^x\right)^2\left(K^y\right)^2/512$}, $\hat{W}_{m,p}=\tau^y_{1}\tau^z_{2}\tau^y_{3}\tau^z_{4}$ (see Fig.~\ref{fig:eff_sq_lattice} for notation of the plaquette sites), and $N$ the number of $K^z$-dimers. Higher orders of the effective model inside the Kitaev layers correspond to multi-plaquette terms.\cite{schmidt08,vidal08} The contributions of the intralayer couplings $J_\perp$ to the effective model can be written as
%
\begin{equation}\label{eq:H_eff_perp}
\mathcal{H}^{0\textrm{QP}}_{\perp,\textrm{eff}} =\sum_{n,m}\mathcal{H}^{(n,m)}_{\perp,\textrm{eff}}\,,
\end{equation}
where $n \in \{1,2,4\}$ and $m \in \{0,2\}$ denotes the order of perturbation in $J_\perp$ and $K_\kappa$ with $\kappa\in\{x,y\}$, respectively, in which the terms appear. These terms are given by
%
\begin{equation}\label{eqn:H_perp_eff}
\begin{split}
\mathcal{H}^{(1,0)}_{\perp,\textrm{eff}}=&\frac{1}{2}J_\perp\sum_{\bi} \tau^z_{1,i}\tau^z_{2,i}~,\\
\mathcal{H}^{(1,2)}_{\perp,\textrm{eff}}=&-\frac{1}{32}J_\perp \sum_{\bi}\sum_{\kappa=x,y}K_{\kappa}^2\tau^z_{1,i}\tau^z_{2,i}~,\\
\mathcal{H}^{(2,0)}_{\perp,\textrm{eff}}=&-\frac{1}{8}J_\perp^2\sum_{\bi}(1+\tau^x_{1,i}\tau^x_{2,i}+\tau^y_{1,i}\tau^y_{2,i}-\tau^z_{1,i}\tau^z_{2,i})~,\\
\mathcal{H}^{(2,2)}_{\perp,\textrm{eff}}=&-\frac{1}{512}J_\perp^2\sum_{\bi}\sum_{\kappa=x,y}K_\kappa^2\Big[ \\
&13-2\,\tau_{1,i}^z\tau_{2,i}^z+10\sum_{\alpha=x,y}\tau_{1,i}^\alpha\tau_{2,i}^\alpha\\
&+5 \Big( \sum_{\alpha=x,y}\tau_{1,i}^\alpha\tau_{2,i}^\alpha\Big)\Big( \sum_{\beta=x,y} \tau_{1,i+n_{\kappa}}^\beta\tau_{2,i+n_{\kappa}}^\beta\Big) \phantom{\sum_{\bi}}\\
&-\Big( \sum_{\alpha=x,y}\tau_{1,i}^\alpha\tau_{2,i}^\alpha\Big) \Big( \tau_{1,i+n_{\kappa}}^z\tau_{2,i+n_{\kappa}}^z \phantom{\Big)} \phantom{\sum_{\bi}}\\
&\phantom{\Big(}+\tau_{1,i-n_{\kappa}}^z\tau_{2,i-n_{\kappa}}^z \Big)+5\,\tau_{1,i}^z\tau_{2,i}^z\tau_{1,i+n_{\kappa}}^z\tau_{2,i+n_{\kappa}}^z\Big]\phantom{\sum_{\bi}},\\
\mathcal{H}^{(4,0)}_{\perp,\textrm{eff}}=&\frac{1}{32}J_\perp^4\sum_{\bi}(1+\tau_{1,i}^x\tau_{2,i}^x+\tau_{1,i}^y\tau_{2,i}^y-\tau_{1,i}^z\tau_{2,i}^z)~.
\end{split}
\end{equation}
%
The decoupled Wen-plaquette models in the limit $J_\perp=0$ are exactly solvable and realize topologically ordered ground states with Abelian anyons as elementary excitations. Next we tackle the question how the intralayer coupling $J_\perp$ destroys this topological order within the effective low-energy description.

\subsection{Duality mapping}

The bilayer Kitaev model with AA stacking exhibits an exact conserved quantity $\hat{\Omega}_p$ for each plaquette $p$, see also Sec. \ref{sec:modelsymm}. As a consequence, also $[\mathcal{H}^{0\textrm{QP}}_{\textrm{eff}},\hat{\Omega}_p]=0$ for all $p$ holds, and the Hilbert space splits in decoupled blocks for each set of eigenvalues $\pm 1$ of the $\hat{\Omega}_p$ operators which can be therefore studied independently. Interestingly, in both limits $J_\perp=0$ as well as $J_\perp\rightarrow\infty$, the exact ground states of the isolated Wen-plaquette models and the product state of singlets on $J_\perp$-bonds belongs to the Hilbert space sector where all eigenvalues of $\hat{\Omega}_p$ operators are $+1$. If there is therefore only a single phase transition between the gapped topologically-ordered phase and the gapped dimer phase, then it has to take place in this sector and can be either a transition of first or second order. If other Hilbert space sectors play a role for the ground-state phase diagram, then these phase transitions between different sectors are definitely first-order phase transitions. One additional reason why these other sectors play most likely no role for the quantum critical behavior of the bilayer Kitaev model, is that elementary excitations of the topologically ordered phase, i.e.~a single eigenvalue $\hat{W}_{m,p}=-1$ on a certain plaquette $p$, as well as single triplons in the dimer phase are exactly localized due to the exact conservation laws. These gapped excitations are therefore very unlikely to close the gap and drive a quantum phase transition. We therefore focus in the following on the sector where all eigenvalues of $\hat{\Omega}_p$ operators are $+1$.

%
\begin{figure}[t]
\begin{center}
\includegraphics[width=\columnwidth]{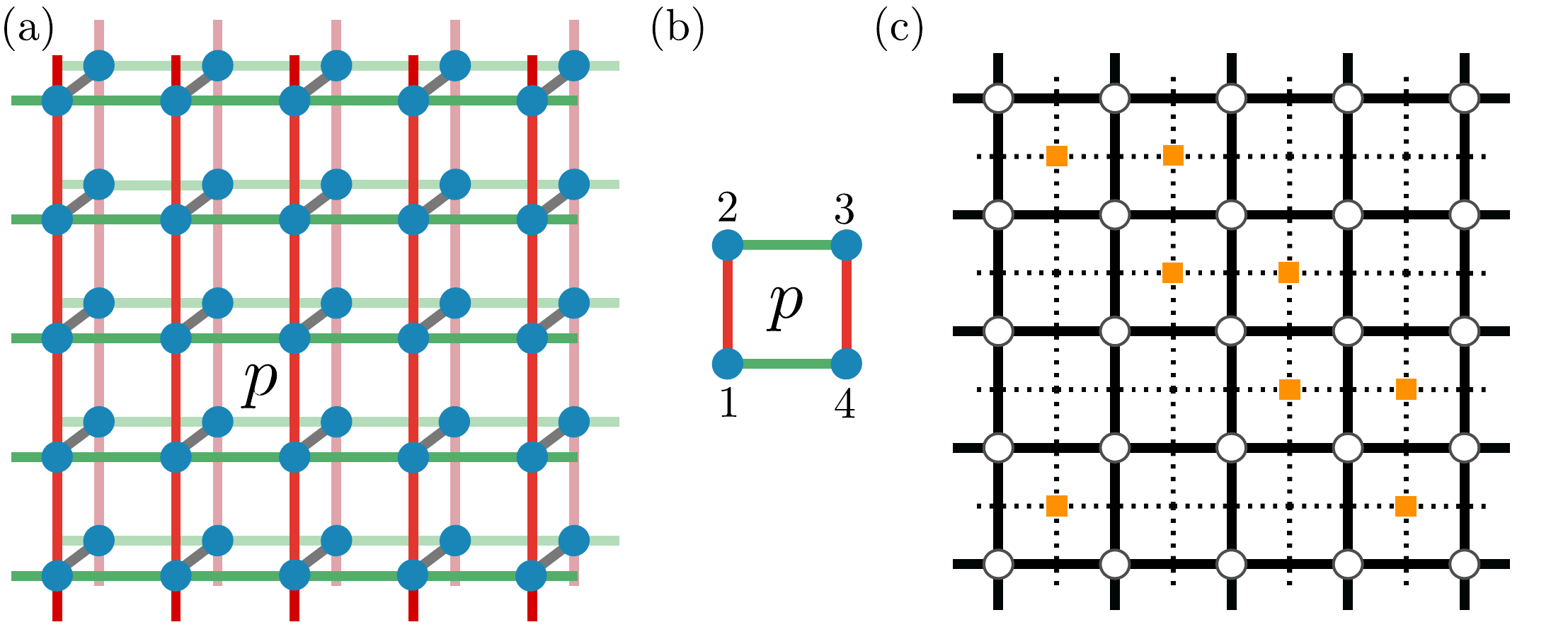}
\caption{(a) Effective bilayer square lattice. (b) Notation of the four sites associated to a plaquette $p$. (c) Effective two-dimensional square lattice. The dual pseudo-spin 1/2 operators act in the centers of the plaquettes. These centers also form a square lattice shown with the dashed lines.
Little orange squares indicate the sites where the sublattice rotation is performed.}
\label{fig:eff_sq_lattice}
\end{center}
\end{figure}
%

In this Hilbert space sector an exact duality mapping is possible by introducing pseudo-spin 1/2 operators $\tilde{\tau}_p^{\alpha}$ centered on plaquettes $p$. Indeed, in order to ensure $\hat{\Omega}_p$ eigenvalue on $p$ to be $+1$, either both $w_p$'s have to be $+1$ or $-1$. This local $\mathbb{Z}_2$  degree of freedom can be represented by the diagonal Pauli matrix $\tilde{\tau}_p^{z}$ in the pseudo-spin bases $|\!\uparrow\rangle$ and $|\!\downarrow\rangle$ for the two combinations. In this pseudo-spin language, the sum of the isolated Wen-plaquette models translates to an effective dual field term
%
\begin{equation}\label{eq:H_eff_Kitaev-field}
\tilde{\mathcal{H}}_{\mathrm{field}}=2E_0-2 C_p\sum_p \tilde{\tau}_p^{z}
\end{equation}
%
so that the topological phase corresponds to a trivial polarized phase with ground state $|\!\uparrow\ldots\uparrow\rangle$ in the dual pseudo-spin language. Flipping a spin costs energy $4C_p$ and represents the elementary gapped excitation in this phase.

The terms proportional to $J_\perp$ introduce quantum fluctations with respect to the field term. Focusing on first- and second-order terms in $J_{\perp}$, the dual expressions read
%
\begin{eqnarray}
\tilde{\mathcal{H}}_{\perp}&=&-\frac{1}{8}J_\perp^2 N +\tilde{J}_{1}^{xx}\sum_{\sqthree}\tilde{\tau}^x_p\tilde{\tau}^x_{p'}\nonumber\\
                           && -\tilde{J}^{xx}_{2}\sum_{\sqtwo}^{\phantom{k}}\tilde{\tau}^x_p\tilde{\tau}^x_{p'} +\tilde{J}_{4}\sum_{\sqfour}\tilde{\tau}^x_{p_1}\tilde{\tau}^x_{p_2}\tilde{\tau}^x_{p_3}\tilde{\tau}^x_{p_4}
\end{eqnarray}
%
with
\begin{equation}
\tilde{J}_{1}^{xx}=\frac{J_\perp}{2} \left[ 1- \frac{1}{16}\left( \left(K^x\right)^2+ \left(K^y\right)^2\right) \right]+\frac{J_\perp^2}{8}
\end{equation}
and $\tilde{J}_4=\tilde{J}_{2}^{xx}=J_\perp^2/8$. The sums are taken over the dark plaquettes of the little pictograms. The dual pseudo-spin operators for the higher-order contributions can also be expressed solely via $\tilde{\tau}^x_{p}$. Finally, we perform the sublattice rotation $\tilde{\tau}^x_{p}\equiv -\tilde{\tau}^x_{p}$, \mbox{$\tilde{\tau}^y_{p}\equiv -\tilde{\tau}^y_{p}$}, and $\tilde{\tau}^z_{p}\equiv\tilde{\tau}^z_{p}$ about the $z$-axis in pseudo-spin space for two consecutive anti-diagonals (see orange squares in Fig.~\ref{fig:eff_sq_lattice}c), which results in the more convenient expression
%
\begin{eqnarray}\label{eq:H_dual}
\tilde{\mathcal{H}}&=& \tilde{\mathcal{H}}_{\mathrm{field}}+ \tilde{\mathcal{H}}_{\perp}\nonumber\\
&=& \tilde{E}_0-\tilde{h}_z\sum_p \tilde{\tau}_p^{z}-\tilde{J}_{1}^{xx}\sum_{\sqthree}\tilde{\tau}^x_p\tilde{\tau}^x_{p'}\nonumber\\
                           && -\tilde{J}^{xx}_{2}\sum_{\sqtwo}^{\phantom{k}}\tilde{\tau}^x_p\tilde{\tau}^x_{p'}-\tilde{J}_{4}\sum_{\sqfour}\tilde{\tau}^x_{p_1}\tilde{\tau}^x_{p_2}\tilde{\tau}^x_{p_3}\tilde{\tau}^x_{p_4}
\end{eqnarray}
%
where $\tilde{E}_0=2E_0-\frac{1}{8}J_\perp^2 N$, $\tilde{h}_z=2C_p$, and all interactions are ferromagnetic.

\subsection{Quantum phase transition for AA stacking}

In this subsection we study \eqref{eq:H_dual} in order to describe the breakdown of the topologically-ordered phase as a function of $J_\perp$, which translates in the dual language to the quantum phase transition out of the polarized phase at large fields $\tilde{h}_z$ and a $\mathbb{Z}_2$ symmetry-broken ferromagnetic phase whenever the interactions are dominant. Interestingly, to leading order in $J_\perp$, the effective model is just a collection of infinitely many decoupled one-dimensional transverse-field Ising chains along one diagonal of the square lattice formed by plaquette centers. The latter can be solved exactly and a second-order quantum phase transition in the 2D-Ising universality class is known to take place at $\tilde{h}_z=\tilde{J}_{1}^{xx}$. This translates to $\left(K^x\right)^2 \left(K^y\right)^2=256 J_\perp$ in the bilayer Kitaev model in units of $K^z=2$. Hence, a tiny coupling $\jp\propto \lambda^4$ closes the gap of the topological phase and induces its breakdown.

The exact dimensional reduction to decoupled one-dimensional systems is destroyed by the second-order contributions in $J_\perp$ and the original two-dimensionality of the bilayer Kitaev model is restored although it stays strongly anisotropic for small $J_\perp$. The order-two interactions $\tilde{J}_{2}^{xx}$ and $\tilde{J}_{4}^{xx}$ both favor a ferromagnetic state. In case $\tilde{J}_{4}^{xx}$ is set to zero, one has two decoupled two-dimensional transverse-field Ising models on anisotropic square lattices. Here the phase transition remains second order and is in the 3D-Ising universality class. In contrast, if only $\tilde{h}_z$ and $\tilde{J}_{4}^{xx}$ are finite (so that both two-spin Ising interactions are zero), then one obtains the Xu-Moore model,\cite{Xu04,Xu05} which itself is isospectral to the compass model \cite{Kugel82,Nussinov05_1,Dorier05} and to the toric code in a transverse field.\cite{Vidal09} All these models possess a self-duality so that the phase transition takes place at $\tilde{h}_z=\tilde{J}_{4}^{xx}$ and is strongly first order. As a conclusion, if all interactions in \eqref{eq:H_dual} are finite, one either has a second-order 3D-Ising transition or a first-order transition.

\begin{figure}[t]
\begin{center}
\includegraphics[width=\columnwidth]{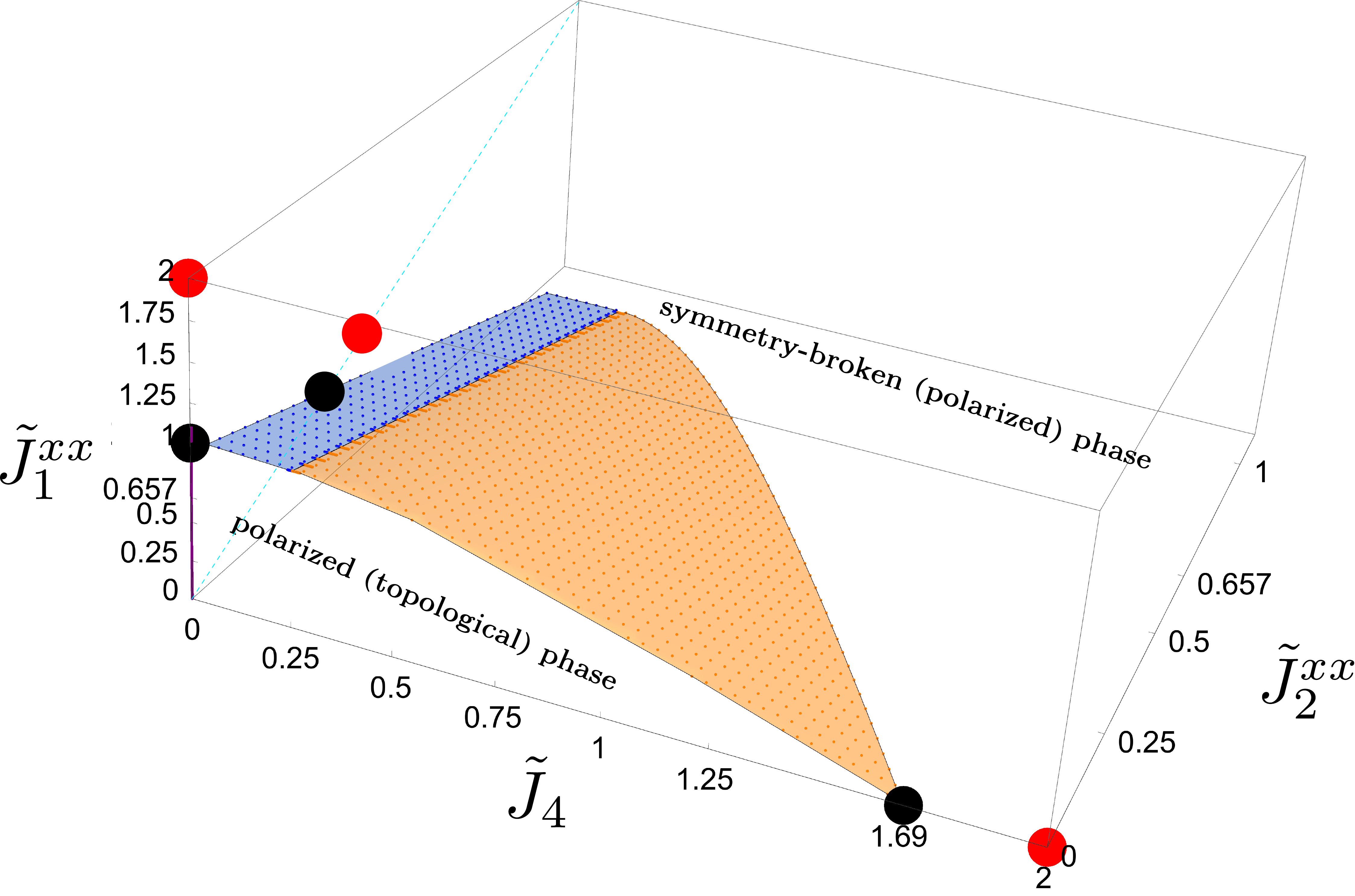}
\caption{The mean-field phase diagram of \eqref{eq:H_dual} as a function of $\tilde{J}_{1}^{xx}$, $\tilde{J}_{2}^{xx}$, and $\tilde{J}_{4}$. The blue (orange) regions are phase transitions of second- (first-)order. The red (black) points indicate the exact (mean-field) critical values for the 1d-TFIM on the $\tilde{J}_{1}^{xx}$-axis, the 2D-TFIM on the dashed cyan line, and the toric code in a transverse field on the $\tilde{J}_{4}$-axis. The purple line is the rescaled physical path generated for $K^x=K^y\approx0.25$, which pierce the phase transition surface at $\tilde{J}_{2}^{xx}\approx \tilde{J}_{4}\approx 0$ and $\tilde{J}_{1}^{xx}\approx 1$. As a consequence, it is the second-order region which is relevant for the quantum phase transition in the anisotropic AA-stacked bilayer Kitaev model.
}
\label{fig:pdf_mf}
\end{center}
\end{figure}

In the following we argue that the quantum phase transition in the bilayer Kitaev model, which corresponds to a specific path in the coupling space of the effective model \eqref{eq:H_dual}, is most likely a 3D-Ising transition. To this end we perform a mean-field calculation by introducing the following one-parameter product state wave function
\begin{equation}\label{eqn:Meanfield1}
\ket{\alpha}=\prod_p\left( \cos(\alpha)\ket{\uparrow}_p+\sin(\alpha)\ket{\downarrow}_p \right)~,
\end{equation}
so that both limiting ground states are taken into account exactly. The polarized phase is realized for $\alpha=0$ and the two ferromagnetic ground states correspond to \mbox{$\alpha=\pm\pi/4$}. The mean-field energy per plaquette of \eqref{eq:H_dual} is then readily calculated and reads
\begin{eqnarray}\label{eq:MF_energy}
  e_{0}^{\rm MF} &=& -\tilde{h}_z \left(\cos^2(\alpha) - \sin^2(\alpha)\right)-16\tilde{J}_{4}\sin^4(\alpha)\cos^4(\alpha)\nonumber\\
               &&-4\big[\tilde{J}_{1}^{xx}+\tilde{J}_{2}^{xx}\big]\sin^2(\alpha)\cos^2(\alpha)\,.
\end{eqnarray}
If one sets $\tilde{h}_z=1$ to fix the overall energy scale, we have located the phase transition between the polarized and the ferromagnetic phase numerically as a function of $\tilde{J}_{1}^{xx}$, $\tilde{J}_{2}^{xx}$, and $\tilde{J}_{4}$. The obtained mean-field phase diagram is plotted in Fig.~\ref{fig:pdf_mf}.

%
%

We stress that this mean-field approach captures the order of the phase transition correctly in all limiting cases, i.e.~second-order phase transitions for the one-dimensional transverse-field Ising chain ($\tilde{J}_1^{xx}$ or $\tilde{J}_2^{xx}$ only), second-order phase transition for the two-dimensional transverse-field Ising model on the square lattice ($\tilde{J}_1^{xx}=\tilde{J}_2^{xx}$ and $\tilde{J}_{4}=0$), and first-order phase transition for pure $\tilde{J}_{4}$. Obviously, the value of the quantum critical points are only correct in a qualitative manner as can be seen when comparing the exact and mean-field results indicated by black and red circles in Fig.~\ref{fig:pdf_mf}.

Most importantly, the quantum phase transition is of second order in a relative wide range of couplings when moving away from the $\tilde{J}_1^{xx}$-axis. Keeping in mind that i) the quantum criticality induced by the first-order contribution in $J_\perp$ takes place exactly on the $\tilde{J}_1^{xx}$-axis for small values of $J_\perp=C_p$ and  ii) higher-order corrections (like the second order) are small for small $J_\perp$, we conclude that the phase transition in the bilayer Kitaev model in the anisotropic limit is most likely a second-order 3D-Ising transition, and we expect the scaling of the critical interlayer coupling to be $\jp\propto \lambda^4$.

\subsection{Quantum phase transition for \sac{} stacking} \label{sec:effmod_ab_ac}

\begin{figure}[t]
\begin{center}
\includegraphics[width=\columnwidth]{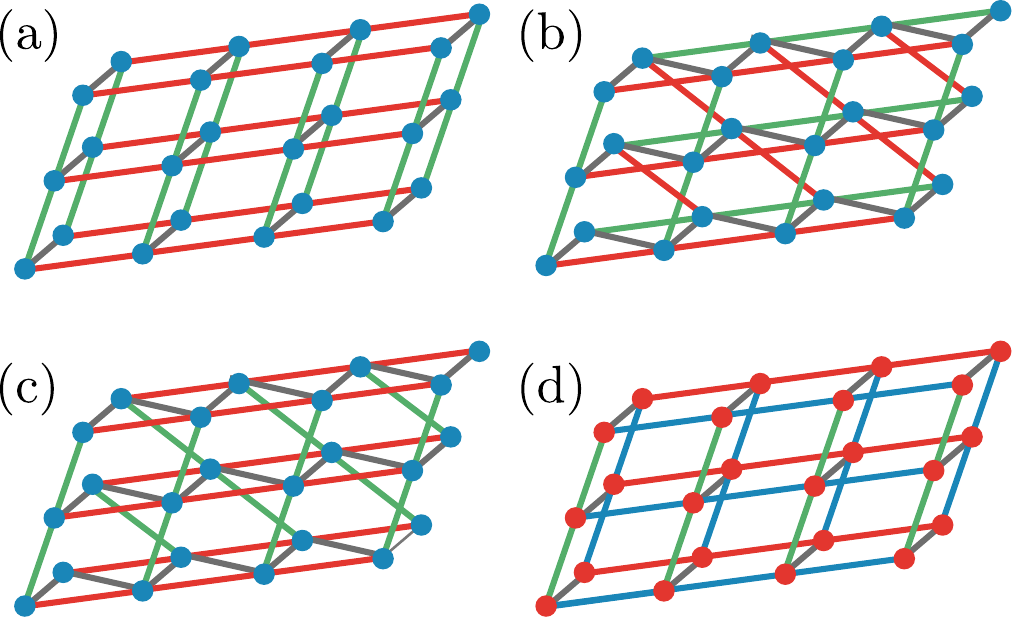}
\caption{
Illustration of the four different anisotropic limits: (a) AA stacking, (b) AB stacking, c) \sac{} stacking when replacing $K^z$-bonds by supersites (shown as filled blue circles) and d) \sbac{} stacking when replacing $K^x$-bonds by supersites (shown as filled red circles). The grey lines represent interlayer $J_\perp$-interactions. Note that the thick grey lines in (a) and (d) refer to the effective interaction from two $J_\perp$-couplings between the supersites. The green, red, and blue lines refer to $K^x$, $K^y$, and $K^z$ interactions in the two Kitaev layers, respectively.
}
\label{fig:stackings_eff}
\end{center}
\end{figure}

We now discuss the transition out of the Kitaev spin liquid in the \sac{} stacking.
The replacement of the strong $K^x$-bonds for the \sac{}-stacked model, the introduction of hardcore bosons and pseudospins, as well as the derivation of the effective low-energy spin model work along the same lines as for the AA stacking.
It is especially the effective network of supersites which is different for the anisotropic limit as illustrated in Fig.~\ref{fig:stackings_eff}.
In addition, the treatment of the corresponding effective low-energy pseudo-spin model is different, since the exact conserved quantities $\hat{\Omega}_p$ for the AA stacking do not exist anymore.
%
%

We therefore use again the pCUT to transform \eqref{eq:ham_v2} into an effective Hamiltonian $H_\mathrm{eff}$ which satifies $[H_\mathrm{eff},\mathcal{Q}]=0$. Obviously, the 0QP effective pseudo-spin model in terms of $\vec{\tau}$ is identical within the two Kitaev layers.
It is only the orientation of effective plaquettes in the two layers which is different for the \sac{} stacking with anisotropy, as illustrated in Fig.~\ref{fig:stackings_eff2}(d).
We therefore again find a Wen-plaquette model up to order four perturbation theory within the layers.
The Wen-plaquette operator $\hat{W}_{m,p}$ in layer $m$ and plaquette $p$ is proportional to $\left(K^y\right)^2\left(K^z\right)^2$ for the limit $K^y,K^z\ll K^x$ in this order.
An essential difference between the different cases is the effective interaction between the layers due to $J_\perp$.
Here we have calculated the two leading orders in $J_\perp$ and fourth orders in $K^y$ and $K^z$ which certainly represent the most important terms as for the AA stacking discussed above.
The effective model can be expressed as
%
\begin{equation}\label{eq:H_eff_stackings2}
\mathcal{H}^{0\textrm{QP}}_{\textrm{eff}}=\sum_{m\in\{1,2\}}\mathcal{H}^{0\textrm{QP}}_{m,\textrm{Kitaev,eff}}+\mathcal{H}^{0\textrm{QP}}_{\perp,\textrm{eff}}~,
\end{equation}
%
where the first term represents the Wen-plaquette model in each Kitaev layer $m$ with plaquette operators $\hat{W}_{m,p}$ as illustrated in Fig.~\ref{fig:stackings_eff2}(d) and the second term is identical to the AA stacking \mbox{$\mathcal{H}^{0\textrm{QP}}_{\perp,\textrm{eff}}= \mathcal{H}^{(1,0)}_{\perp,\textrm{eff}}+\mathcal{H}^{(1,0)}_{\perp,\textrm{eff}}$}, since again two $J_\perp$-couplings connect neighboring supersites.
As a consequence, this limit of the \sac{} stacking behaves similar to the AA stacking as already seen in the mean-field treatment.
One has a trivial phase for $K^y,K^z\ll J_\perp$, which is adiabatically connected to isolated $J_\perp$-dimers for $K^y$=$K^z$=$0$ (see Fig.~\ref{fig:stackings_eff}(d)), and a topological phase for the other limit of weakly coupled Kitaev layers.
It is reasonable that the quantum phase transition between both phases is, as discussed for the AA stacking, either second-order in the (2+1)D Ising universality class or of first-order.
The main difference to the AA stacking is that the Wen-plaquette operators are not the same type on opposite plaquettes of the two layers (see Fig.~\ref{fig:stackings_eff2}(d)), since in the current case the $K^y$ and $K^z$ couplings are rotated by 90$^\circ$ from one Kitaev layer to the other.
As a result, there exist no exact conserved quantities $\hat{\Omega}_{p}$ and the effect of $\mathcal{H}^{0\textrm{QP}}_{\perp,\textrm{eff}}$ on the excitations of the topological phase is different.
In leading order in $J_\perp$, there are almost no mobile excitations (plaquettes with $\omega_{m,p}=-1$) at all, e.g.~single excitations on one of the two Kitaev layers are not allowed to hop. One exception are two excitations located on the different Kitaev layers as close as possible, but not exactly on top of each other, which are able to move but only in one dimension.
Altogether, the constraint mobility of the excitations in the topological phase point towards a first-order phase transition in the \sac{} stacking for $K^y,K^z\ll K^x$ similarly to the toric code in a transverse field.\cite{Vidal09}


%
%

\section{Effective chain models and macro-spin phases in the AB and \sbac{} stackings}
\label{sec:novel}

Considering the discussion in Secs. \ref{sec:mf} and \ref{sec:series_aniso}, it has become evident that at strong anisotropies the AB- and \sbac{}-stacked models result in a striking geometry of chains consisting of the strong bonds (as illustrated in Fig. \ref{fig:chains_illu}), with weak residual interactions between them.
In the MMFT, this geometry resulted in an effective one-dimensional dispersion for the itinerant Majorana fermions. Similarly, the series expansion (based on the limit $\jp \ll K$) features triplons dispersing along these chains.

The purpose of this section is to study the consequences of this particular geometry.
Our approach is twofold: we first consider the case of $K \gg \jp$ and $\lambda \ll 1$ and construct an effective Ising model for pseudospins formed from $K^z$-dimers.
Secondly, in order to study the transition from \pddim{} to the \pdchains{} phase,  we consider the case $K \ll \jp$ and successive triplon condensation, obtaining a transverse-field Ising chain (TFIC) as an effective model in the low-energy subspace spanned by interlayer-dimer singlet and triplet states.
The symmetry-broken phase of the TFIC corresponds to the \pdchains{} phases, with the ground state corresponding to an essentially classical macro-spin.
Finally, we discuss possible interactions between these macro-spins in the respective stackings.

\subsection{Effective model for Kitaev dimers at $K \gg \jp$.}

\begin{figure}[t]
\begin{center}
\includegraphics[width=0.7\columnwidth]{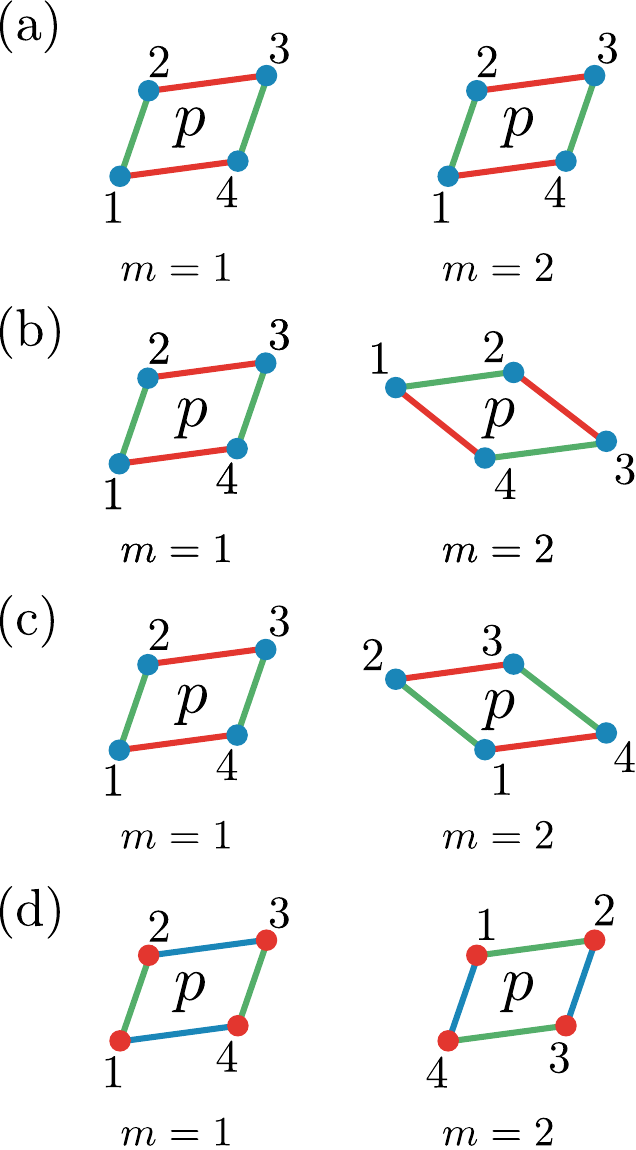}
\caption{Illustration of the plaquette operators $\hat{W}_{m,p}$ for Kitaev layer $m=1$ (left) and $m=2$ (right) for the different anisotropic limits: (a) AA stacking, (b) AB stacking, (c) \sbac{}-stacking when replacing $K^z$-bonds by supersites (shown as filled blue circles) and (d) \sac{}-stacking when replacing $K^x$-bonds by supersites (shown as filled red circles). The numbering 1-4 of the plaquette sites refers to the definition of the Wen-plaquette operator \mbox{$\hat{W}_{p}=\tau^y_{1}\tau^z_{2}\tau^y_{3}\tau^z_{4}$} in all cases.}
\label{fig:stackings_eff2}
\end{center}
\end{figure}

The effective geometries for the AB- and \sbac{}-stacked models in the limit $K^x,K^y\ll K^z$ are shown in Fig.~\ref{fig:stackings_eff}(b,c).
Up to order four in $K^x$, $K^y$, and second order in $J_\perp$ we find for these two cases
%
\begin{equation}\label{eq:H_eff_stackings}
\mathcal{H}^{0\textrm{QP}}_{\textrm{eff}}=\sum_{m\in\{1,2\}}\mathcal{H}^{0\textrm{QP}}_{m,\textrm{Kitaev,eff}}+\mathcal{H}^{0\textrm{QP}}_{\perp,\textrm{eff}}~,
\end{equation}
%
where the first term represents the well-known Wen-plaquette model in each Kitaev layer $m$ as in Eq.~\eqref{eq:H_eff_Kitaev}, where, however, the notation of the plaquette sites in $\hat{W}_{p}=\tau^y_{1}\tau^z_{2}\tau^y_{3}\tau^z_{4}$ depends on the stacking and on the layer $m$ as illustrated in Fig.~\ref{fig:stackings_eff2}(b,c).

%

The contributions of the intralayer couplings $J_\perp$ to the effective model can be written in the simple form
%
\begin{eqnarray}\label{eq:H_eff_perp_stackings}
  \mathcal{H}^{0\textrm{QP}}_{\perp,\textrm{eff}} &=& -\frac{J_\perp^2}{16} N_\perp + \left( \frac{1}{4}J_\perp+\frac{1}{16}J_\perp^2\right)\nonumber\\
                                           && \sum_{\bi} \left( \tau^z_{1,i}\tau^z_{2,i+\frac{\delta_x}{2}} + \tau^z_{2,i+\frac{\delta_x}{2}}\tau^z_{1,i+\delta_x}\right)~,
\end{eqnarray}
where the sum runs over all supersites of one Kitaev layer (filled blue circles in Fig.~\ref{fig:stackings_eff}(b,c)).

One important difference compared to the AA stacking is the macro-spin phase triggered by the effective Ising interactions in Eq.~\eqref{eq:H_eff_perp_stackings} due to $J_\perp$.
Indeed, for $K^x$=$K^y$=$0$, one has isolated Ising chains in both cases (Fig.~\ref{fig:stackings_eff2}b-c), so that a sub-extensive degeneracy $2^{N_{\rm c}}$ with $N_{\rm c}$ number of Ising chains arises due to the two exact ground states $|\uparrow\downarrow\uparrow\downarrow\ldots\rangle$ and $|\downarrow\uparrow\downarrow\uparrow\ldots\rangle$ of each Ising chain in this limit.
Considering the Wen-plaquette operator, which connects neighboring Ising chains, as a perturbation on this degenerate manifold, no effective interaction between the ground state arises up to order eight perturbation theory in $K^x,K^y$, since acting with $\hat{W}_{m,p}$ on the same plaquette $p$ leads to the same energy reduction of each degenerate ground state.
Altogether, there is no obvious perturbative mechanism to lift this degeneracy in within this effective model.
In the opposite limit $J_\perp\ll K^x,K^y \ll K^z$ in Eq.~\eqref{eq:H_eff_stackings}, as for the AA stacking, one has a gapped topological phase for the weakly coupled Kitaev layers.
Consequently, there must be also a quantum phase transition between this topological and the macro-spin phase discussed before, which is most likely of first-order nature.


\subsection{Effective model for interlayer dimers at $\jp \gg K$}

We now approach the \pdchains{} phases from the dimer phase, $K^z \ll \jp$, by first discussing a $K^z$-$J_\perp$ Ising-Heisenberg chain which is formed at $\lambda =0$.
This chain can be mapped exactly on a transverse-field Ising chain so that its quantum phase diagram is known exactly as a function of $K^z/J_\perp$. Indeed, if we use the four states $\ket{s}$ and $\ket{t_\alpha}$ with $\alpha\in\{x,y,z\}$ of Heisenberg dimers as a basis to describe the Ising-Heisenberg chain, then it can be readily seen that Ising interaction between dimers only affect the states $\ket{s}$ and $\ket{t_z}$ while the other two triplet states are not affected at all.
We therefore can introduce a pseudo-spin 1/2 on each Heisenberg dimer by identifying $\ket{\downarrow}\equiv \ket{s}$ and $\ket{\uparrow}\equiv \ket{t_z}$. In terms of pseudo-spin-$1/2$ Pauli matrices $\tilde{\tau}^\alpha$ with $\alpha\in\{x,y,z\}$, the Heisenberg interaction then becomes, up to an irrelevant constant, an effective field term $J_\perp / 2 \sum_{d} \tilde{\tau}_d^z$ where the sum runs over all dimers $d$.
The intra-dimer Ising interaction always flips the pseudo-spin state on two adjacent dimers. As a consequence, it corresponds also to an (effective) Ising interaction in terms of pseudo-spins and reads $(K^z/4)\sum_{\langle d,d'\rangle} \tilde{\tau}_d^x\tilde{\tau}_{d'}^x$. In total, this gives an effective transverse-field Ising chain
\begin{equation}
  \mathcal{H}_{\rm c}= \frac{J_\perp}{2} \sum_{d} \tilde{\tau}_d^z - \frac{K^z}{4}\sum_{\langle d,d'\rangle} \tilde{\tau}_d^x\tilde{\tau}_{d'}^x\quad ,
\end{equation}
which is known to realize a continuous quantum phase transition in the 2D Ising universality class for \mbox{$2J_\perp=\pm K^z$}.

Coming back to the full bilayer Kitaev model for $\lambda=0$, we have a collection of decoupled Ising-Heisenberg chains where each TFIC possesses a quantum phase transition at \mbox{$2J_\perp=K^z$}.
 This is also evident from the series expansion (cf.~Sec.\ref{sec:series_aniso}), in which the triplon gap closes at \mbox{$\jp / K^z = 0.5$}, also in the presence of finite $\lambda$. Note that the MMFT shows the critical $\jp /K^z \simeq 0.4$ and is thus also close to the exact value.

For $2J_\perp>K^z$, each chain has a unique gapped ground state which is adiabatically connectected to the product state of singlets $\ket{s}\cdots\ket{s}$, $\ket{\downarrow}\cdots\ket{\downarrow}$ in pseudo-spin language, being the ground state for $K^z=0$. Obviously, coupling the chains $\lambda\neq 0$ in this parameter regime, one still has a unique ground state corresponding to the featureless dimer paramagnet.

However, the situation is different for $2 J_\perp<K^z$. Then, for $\lambda=0$, each Ising-Heisenberg chain is in one of the two ground states of the symmetry-broken phase and there is a degenerate manifold of $2^{N_{\rm c}}$ states with $N_{\rm c}$ the number of chains.
Note that the individual chain ground states are adiabatically connected to the Ising ground states $\ket{\Rightarrow} \equiv |\rightarrow\cdots\rightarrow\rangle$ and $\ket{\Leftarrow} \equiv \ket{\leftarrow\cdots\leftarrow}$ for $J_\perp=0$. The chain states $\ket{\Rightarrow}$ and $\ket{\Leftarrow}$ can therefore be interpreted as the two orientations of a large macro-spin.
We however emphasize that in terms of the microscopic Kitaev model, the respective macro-spin ground states correspond to antiferromagnetic configurations of the local moments.
The full bilayer Kitaev model for $\lambda=0$ and $2J_\perp<K^z$ is then effectively a chain of decoupled macro-spins.

\subsection{Macro-spin interactions and classical spin liquid}

The final question is what kind of effective interaction between the macro-spins is introduced for finite $\lambda$ and whether or not this interaction leads to a unique ground state.

In the series-expansion treatment in Sec.~\ref{sec:series_aniso}, it was found that the triplon gap closes at $\vec k = 0$ in the AB stacking, corresponding to a ferromagnetic macro-spin interaction (yielding an antiferromagnetically ordered state for the local moments).
We call this phase \pdmacaf{} in Fig. \ref{fig:header}.

In the \sbac{}-stacked model, however, the triplon gap closes along a line in momentum space, which is consistent with a macroscopic degeneracy between the macro-spins.
We thus deduce that this phase realizes a classical spin liquid, dubbed \pdmacsl{}, formed of macroscopically large spins with no residual interaction.

We complement the results from the series expansion with analytical arguments by peturbatively integrating out the microscopic $K^x$- and $K^y$-interactions to (possibly) obtain an effective interaction for the macro-spins $\ket{\Leftarrow}$ and $\ket{\Rightarrow}$.

\begin{figure}[!tb]
  \includegraphics[width=\columnwidth]{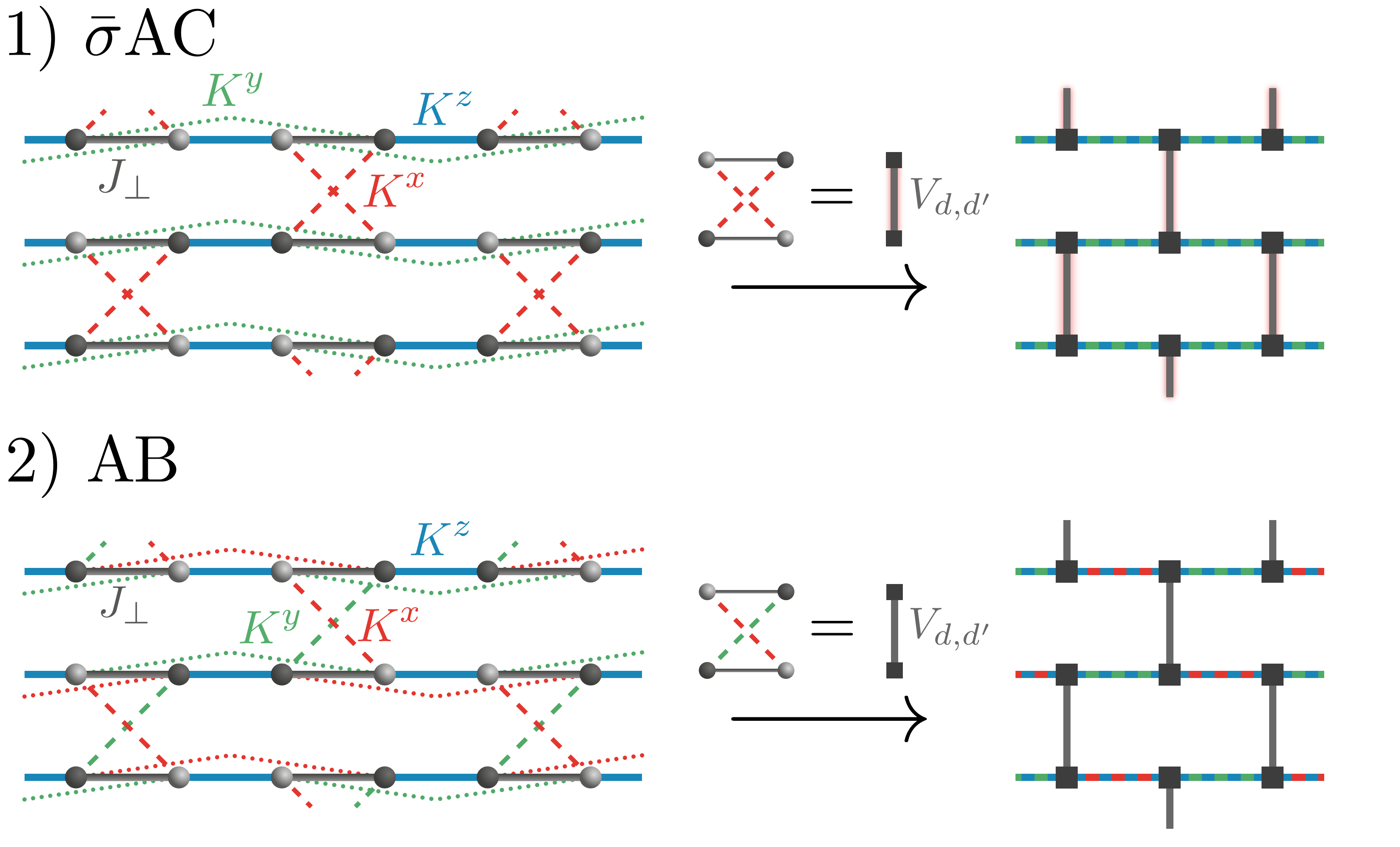}
  \caption{Effective chains consisting of dimers coupled by $K^z$.  Effective brick-wall models are obtained by replacing each dimer by a pseudospin. 1) For the \sbac{} stacking, there is an intra-chain coupling $K^y$, and the chains are coupled via $V_{d,d'}$ which consists of $K^x$ couplings. 2) In the case of AB stacking, both inter- and intra-chain couplings are due to $K^x$ and $K^y$-interactions.}
  \label{fig:eff_chains}
\end{figure}

\paragraph{\sbac{} stacking.} A schematic model for the chains and inter- and intra-chain interactions is shown in Fig.~\ref{fig:eff_chains}.
Since the $K^y$-interaction acts within each chain, it does not affect the degeneracy of the macro-spins.
The perturbation $V$ however, which couples the effective chains (described by the states $\ket{\Rightarrow},\ket{\Leftarrow}$), acts solely on the $x$-components of the microscopic local moments.
For simplicity, we consider the action of $V$ on two isolated \mbox{dimers $d,d'$},
\begin{equation}
  V_{d,d'} = - K^x \left( S^x_{d,0} S^x_{d',1} + S^x_{d,1} S^x_{d',0} \right),
\end{equation}
where $0,1$ describe the two positions within each dimer.
As described in Sec.~\ref{sec:modelsymm}, all stackings possess a $C^\ast_\alpha$-symmetry of rotating all spins by $\pi$ around the $\alpha$-axis.
Now perform  $(C^\ast_x) : (S^x,S^y,S^z) \to (S^x,-S^y,-S^z)$ on every second chain, denoting this operation $U$.
It is clear that $V_{d,d'}$ and thus also $V$ is symmetric under this symmetry operation (as is $\mathcal H_c$), while the macro-spin orientation is reversed, i.e. $U\ket{\Leftarrow} =  \ket{\Rightarrow}$.
We thus find that for all powers $n$ the matrix elements of $V^n$ between neighboring macro-spins fulfill
\begin{equation}
  \braket{\Rightarrow \Rightarrow | V^n |\Rightarrow \Rightarrow} = \braket{\Rightarrow \Leftarrow | V^n |\Rightarrow \Leftarrow},
\end{equation}
such that parallel and antiparallel macro-spin configurations remain degenerate to all orders in perturbation theory in $K^x$.
These considerations are consistent with the fact that in the series expansion the gap of the triplet dispersion closes along a line in momentum space (cf. Sec.~\ref{sec:series_aniso}) in the anisotropic limit.
The \pdchains{}-phase in the AC stacking thus realizes a phase with macroscopic degeneracy, dubbed ``Macro-spin liquid'' (MAC-L).

\paragraph{AB stacking.} For the AB-stacked model, we consider the perturbation $V$ to act on the product states built from isolated dimers $d$ and $d'$, which now reads
\begin{equation}
  V_{d,d'} = - \left(K^x S^x_{d,0} S^y_{d',1} + K^y S^y_{d,1} S^x_{d',0} \right),
\end{equation}
with $K^x = K^y = \lambda K^z$ and $\lambda \ll 1$.
For an effective Hamiltonian $\mathcal{H}^\mathrm{eff}$ to lift the degeneracy between the chains, we require that  $\Delta_{\mathrm{Lift}} = \braket{\Rightarrow \Rightarrow | \mathcal{H}^\mathrm{eff} |\Rightarrow \Rightarrow} - \braket{\Rightarrow \Leftarrow | \mathcal{H}^\mathrm{eff} |\Rightarrow \Leftarrow} \neq 0$.
Again considering the action of $V$ on two isolated dimers, we find that
\begin{equation}
  \Delta_{\mathrm{Lift}} \propto \pm \left( \braket{s_{d} t^z_{d'} | \mathcal{H}^\mathrm{eff} | t^z_{d} s_{d'}} + \braket{s_{d} s_{d'} | \mathcal{H}^\mathrm{eff} | t^z_{d} t^z_{d'}} \right) \label{eq:deltalift}
\end{equation}
in the singlet-triplet basis for two local dimers, comprising the matrix elements for the transfer of a single triplet between two chains, and the creation/annihilation of a triplets on both dimers.
In perturbation theory we find that these two processes cancel to all orders considered by us: the matrix element for the transfer of a triplet involve intermediate states with mixed triplet flavors (such as $\sim \ket{t^x_{d} t^y_{d'}}$) which have a complex overlap with the initial state $\ket{t^z_{d} s_{d'}}$.
The matrix element between any mixed intermediate state and the flipped dimer $\ket{s_{d} t^z_{d'}}$ has an opposing complex phase, leading to an overall positive sign for the transfer process (this argument can be iterated in higher orders of perturbation theory, as the mixed $\ket{t^x t^y}$ and $\ket{t^y t^x}$ remain the only accessible intermediate states), while the second term in Eq.~\eqref{eq:deltalift} carries a negative sign. Since the energies for the excited intermediate states are equal, we find that $\Delta_{\mathrm{Lift}} = 0$.
Effects due to intra-chain interactions (which is essentially a third-nearest neighbor coupling) do not give rise to new intermediate states which would alter above considerations.

These perturbative arguments, combined with the information from the series expansion which signals a ferromagnetic interaction between the macro-spins, suggest that the corresponding bulk energy gain is non-analytic (likely exponential) in $\lambda$. This is not in contradiction with the relevant transverse piece of the triplon dispersion scaling as $\lambda^2$ because the interaction of macro-spins involves an infinite number of single-particle excitations.
Hence, the AB stacking at anisotropies realizes a gapped antiferromagnet (since each macro-spin corresponds to an antiferromagnetic ordering of the local moments), dubbed MAC-AF.

\section{Stability of \pdksli{} and interlayer-coherent $\pi$-flux phase}
\label{sec:pert_ksl}

\subsection{Perturbation theory in $\jp$}

The purpose of this section is to study the stability of the \pdksli{} phase. We argue that, starting from two decoupled Kitaev spin liquids in both layers and coupling them perturbatively (i.e. $\jp \ll K$), there can not be a gap opening in the spectrum of the itinerant Majorana mode.
We consider low-energy processes which are below the flux gap of the Kitaev model, and therefore leave each layer in the flux-free ground states.
It is clear that the only term which directly influences the spectrum involves a matter-Majorana in each layer, $\sim \chi^0_{1i} \chi^0_{2i}$.
Considering $\mathcal{H}_\perp$ which acts with a local spin operator $S^\alpha_j = \ii \chi^0_j \chi^\alpha_j$ (adding a matter-Majorana and creating a flux pair adjacent to the $\alpha$-bond emanating site $j$) in each layer, it is clear that such a process would necessarily also change the number of flux excitations in each layer, and therefore would not stay in the flux-free sector.
These explicit arguments are consistent with the fact that the perturbation at hand is time-reversal symmetric, and the gaplessness of the Kitaev spin liquid is protected against small time-reversal symmetric perturbations.\cite{kitaev06}

Indeed, it has been argued that a generic lowest-order inter-layer transport process transfers pairs of spinons between the layers as these fractionalized excitations are non-local in nature and thus result in vanishing matrix elements for single-spinon hopping processes,\cite{wcmb17} such as $\vec S_{1i} \cdot \vec S_{2i}$ in the present case. In this case, such a process would correspond to Majorana pair hopping.\cite{syb16}

To study the effects of such a pair hopping term, we expand the matter Majorana $\chi^0$ around the Dirac nodes and obtain an effective $(2+1)$-dimensional action for a free fermion $\psi$.
Power counting yields $[\psi]=1$ and thus a four-fermion process which would correspond to Majorana pair hopping between the layers has $[\bar \psi \bar \psi \psi \psi] = 4$, and is therefore irrelevant in $(2+1)$ dimensions (as are even higher-order processes), such that the \pdksli{} phase is stable for small $\jp /K^z$.

\subsection{Spontaneous interlayer coherence}

Within our mean-field treatment, Sec.~\ref{sec:mf}, we do find a transition to a state which, on the one hand, still possesses several features of the Kitaev spin liquid and, on the other hand, has non-vanishing amplitudes for inter-layer hopping of the Majorana fermions.
This state thus resembles the interlayer-coherent phases discussed previously for quantum Hall bilayer systems.\cite{moon95}
We emphasize that, given that all interlayer-transport operators are irrelevant, this phase has to occur spontaneously, in similarity to exciton condensate phases in electron--hole, and equivalently, electron--electron bilayer systems.\cite{emcd04}

In the MMFT, we observed that it is crucial for the gap to occur that the Kitaev mean fields in the two layer occur with opposite signs $u^{0,\alpha}_0 = -u^{0,\alpha}_1$, which can be understood by considering the symmetry properties of the dispersing Majorana mode.
We note that the Dirac nodes of the dispersing $\chi^0$-band of the Kitaev spin liquid are protected against (small) perturbations by combined time-reversal $\mathcal T$ and inversion symmetry $\mathcal I$.
In the bilayer system at hand, a global inversion symmetry also interchanges the layer index $m$, i.e. $\mathcal I:(x,y,m) \to (-x,-y,-m)$.
Spoiling inversion symmetry by choosing opposing signs for $u^{0,\alpha}$ in the two layers, $\mathcal T \mathcal I$ is no longer a symmetry which protects the nodal points, and a gap is allowed to open upon applying a perturbation.
This result is also easily obtained in the $\Ztwo$-gauge theory description, by performing a gauge transformation in one of the two layers which flips the sign of the gauge field by applying the $D = 4 \chi^0 \chi^1 \chi^2 \chi^3$-operator on only one sublattice. This operation reverses the sign for the inter-layer hopping for the itinerant Majoranas on those dimers connecting sites of this particular sublattice, leading to a staggered hopping between the two layers.
The itinerant Majorana $\chi^0$ thus picks up a $\Ztwo$ phase of $-1$ when going around the plaquette $P=S^\alpha_{1,A} S^\alpha_{1,B} S^\alpha_{2,B} S^\alpha_{2,A}$ (with a fixed, but arbitrary $\alpha=x,y,z$).
These inter-layer plaquettes thus contain a $\pi$-flux, in resemblance of flux states previously discussed for single-layer spin liquids.\cite{maa88,maa89}

On a mean-field level, the cumulant $\kappa(P)$ of the four-spin plaquette can be seen to be yield an order parameter for this flux phase, as
\begin{subequations}
\begin{align}
    \langle \kappa(P) \rangle &= \langle S^\alpha_{1,A} S^\alpha_{1,B} S^\alpha_{2,B} S^\alpha_{2,A} \rangle - \langle S^\alpha_{1,A} S^\alpha_{1,B} \rangle \langle S^\alpha_{2,A} S^\alpha_{2,B} \rangle \nonumber \\
    &\qquad- \langle S^\alpha_{1,A} S^\alpha_{2,A} \rangle \langle S^\alpha_{1,B} S^\alpha_{2,B} \rangle \\
    &= -w^0_A w^0_B  u_1^\alpha u_2^\alpha - u^0_1 u^0_2 w^\alpha_A w^\alpha_B,
\end{align}
\end{subequations}
where we have used that $\langle S_{1,A}^\alpha S_{2,B}^\alpha \rangle = \langle S_{1,B}^\alpha S_{2,A}^\alpha \rangle = 0$.
Within our mean-field decoupling, it is thus evident that $\kappa(P)$ is only finite if both $w$ and $u$ are finite, and $\kappa(P)$ is sensitive to a $\pi$-flux in the plaquette (which can be described, as above, by a staggered inter-layer hopping with uniform $u_1 = u_2$, or equivalently by an antisymmetry $u_1 = - u_2$ and uniform inter-layer hopping). However, the utility of $\kappa(P)$ as an order parameter for the flux phase beyond mean-field theory is unclear.

Importantly, the opening of a gap in the spectrum of the Majorana mode in the model at hand can only be achieved by introducing a $\pi$-flux to the inter-layer plaquettes. As the dispersion of the itinerant Majorana fermion directly influences thermodynamic and response functions, the gapping of the systems can be used as a diagnostic for the occurrence of FLUX, in analogy to previous studies of flux phases.\cite{maa89}


\section{Conclusion and outlook}

We have presented a comprehensive study of bilayer Kitaev models that differ in the stacking pattern of the Kitaev bonds.
At small $\jp / K$, these models exhibit a $\Ztwo$-fractionalized spin liquid phase described by the Kitaev model in each layer. We have studied the breakdown of this topological phase and the transition to the dimer paramagnet by deriving effective models in the anisotropic limit.
Additionally, two further stacking variants of the model lead to novel macro-spin phases\cite{kaul10,sikkenk16} at finite $\jp / K$ and strong anisotropies, which can be described in terms of macro-spins emerging from interlayer chains.
These macro-spins can be either coupled ferromagnetically (realizing a microscopic antiferromagnet) or remain degenerate and thus constitute a classical spin liquid.
Moreover, we have discussed the possibility of a flux phase with spontaneous interlayer coherence to occur in bilayer spin-liquid systems.

We have made use of complementary methods in order to study all regions of the phase diagrams for the problem at hand: While the Majorana mean-field theory is exact in the limit $K \gg \jp$, series expansion techniques allow for a controlled study of the dimer phase (for $K \ll \jp$). Effective models for the anisotropic limit allow for further insight into critical properties of the model at hand.
Whenever the respective methods can be expected to yield reliable results in the same parameter regime, a comparison shows overall consistency:
The result obtained through MMFT for the critical $\jp$ for the MAC-DIM transition is in good agreement with $\jp = 0.5 K$ as obtained from both series expansion and an effective model. Moreover, both MMFT and series expansion techniques yield a vertical shape for the transition line when considering finite inter-chain couplings.

With recent advances in numerical methods, most notably iDMRG,\cite{cullo08,kzmbp13} reliable quantitative studies of the bilayer Kitaev model are in principle possible and can be expected to yield further insight into phases and critical properties of the model.

Several materials with dominant Kitaev interactions have been identified in recent years, most notably $\alpha$-RuCl$_3$,\cite{sears15,ziad16,ban15} with an effectively layered crystal structure.\cite{johns15} While synthesis of honeycomb monolayers and subsequent re-stacking has been reported,\cite{web16} engineering an inter-layer Heisenberg interaction would be an interesting avenue for future experimental efforts.

Our study has shown that the bilayer Kitaev model shows an exciting phenomenology with several unexpected novel phases. We hence believe that bilayer spin liquids and their critical phenomena constitute a rich and promising field for future studies.


\acknowledgments

We thank M. Garst and T. Meng for useful discussions. This research has been
supported in part by the DFG via SFB 1143 (project A02). W.B. acknowledges partial
support by QUANOMET, CiNNds, and kind hospitality of the PSM, Dresden.


\bigskip

\textit{Note added.} Upon completion of this paper, we became aware of parallel work on the bilayer Kitaev model: Ref.~\onlinecite{koga18} exclusively considered the AA-stacked bilayer model in the isotropic case $\lambda = 1$, with results which are largely consistent with ours. We note that they conclude the transition to be first order whereas our results appear more consistent with second order.


\appendix

\section{Bond-operator theory} \label{sec:BOT}

A simple and efficient description of the large-$\jp$ dimerized phase (\pddim{}) is given by bond-operator theory, \cite{bondop} where the spin-1 excitation (triplons) are treated as auxiliary bosons with a hard-core constraint.
With $|t_0\rangle = [|\uparrow\downarrow\rangle - |\downarrow\uparrow\rangle]/\sqrt{2}$ being the singlet state,
while $|t_x\rangle = -[|\uparrow\uparrow\rangle - |\downarrow\downarrow\rangle]/\sqrt{2}$, $|t_y\rangle = \ii[|\uparrow\uparrow\rangle + |\downarrow\downarrow\rangle]/\sqrt{2}$, and
$|t_z\rangle = [|\uparrow\downarrow\rangle + |\downarrow\uparrow\rangle]/\sqrt{2}$ the  spin-$1$ triplet states, the triplon operators are defined as
$t_{\gamma}^{\dagger} |t_0\rangle = |t_\gamma \rangle$ ($\gamma=x,y,z$).
Note that the bond-operator theory can be generalized to magnetically ordered phases as well.\cite{sommer}

In terms of the triplon operators, the spin operators on each layer are represented as follows:
\begin{equation}
\label{eq:spin_tr}
{S}_{i1,2}^\alpha = \frac{1}{2} \left(
  \pm t_{i\alpha}^\dagger P_i
  \pm P_i t_{i\alpha}
  - \ii\epsilon_{\alpha\beta\gamma} t_{i\beta}^\dagger t_{i\gamma} \right) \,,
\end{equation}
where $P_i = 1- \sum_{\gamma=1}^{3} t_{i\gamma}^\dagger t_{i\gamma}$ is the projection operator to handle the constraint \cite{collins} of physical Hilbert space. Inserting the above
expressions in the bilayer Kitaev model, Eqs. \eqref{eq:haa}-\eqref{eq:hac}, one obtains an interacting triplon Hamiltonian. Expanding in the number of triplon operators, the leading term is $\mathcal{H}_0 = - 3\jp/4 N$, and the bilinear piece reads
\begin{align}
\label{eq:h2_bo}
\mathcal{H}_{ha}^{\mu\nu\delta} &= J \sum_{i \alpha} t_{i\alpha}^{\dagger} t_{i\alpha}  \nonumber \\
&-\sum_{\langle ij \rangle_1} \bigg[ \frac{K^x}{4} \left( t_{ix}^{\dagger} t_{jx}^{\dagger} + t_{ix}^{\dagger} t_{jx} + H.c. \right)  \nonumber \\
&~~~~~~~~~~+ \frac{K^{\mu}}{4} \left( t_{i\mu}^{\dagger} t_{j\mu}^{\dagger} + t_{i\mu}^{\dagger} t_{j\mu} + H.c. \right) \bigg]   \nonumber \\
&-\sum_{\langle ij \rangle_2} \bigg[ \frac{K^y}{4} \left( t_{iy}^{\dagger} t_{jy}^{\dagger} + t_{iy}^{\dagger} t_{jy} + H.c. \right)   \nonumber \\
&~~~~~~~~~~+ \frac{K^{\nu}}{4} \left( t_{i\nu}^{\dagger} t_{j\nu}^{\dagger} + t_{i\nu}^{\dagger} t_{j\nu} + H.c. \right) \bigg]  \nonumber \\
&-\sum_{\langle ij \rangle_3} \bigg[ \frac{K^z}{4} \left( t_{iz}^{\dagger} t_{jz}^{\dagger} + t_{iz}^{\dagger} t_{jz} + H.c. \right)  \nonumber \\
&~~~~~~~~~~+ \frac{K^{\delta}}{4} \left( t_{i\delta}^{\dagger} t_{j\delta}^{\dagger} + t_{i\delta}^{\dagger} t_{j\delta} + H.c. \right) \bigg] \,,
\end{align}
where $N$ is the number of dimer sites, and $\mu \nu \delta=x,y,z$ (or permutations) denote the Kitaev couplings in layer 2 according to the chosen stacking, see Sec.~\ref{sec:model}. 
Fourier transforming the triplon operators yields a momentum-space representation of the bilinear Hamiltonian as:
\begin{equation}
\label{eq:h2k_bo}
\mathcal{H}_{ha,\kk}^{\mu\nu\delta} = \frac{1}{2} \sum_{\kk, \alpha} \Psi_{\kk,\alpha}^{\dagger} \mathcal{M}_{\kk,\alpha} \Psi_{\kk,\alpha} \,,
\end{equation}
where $\Psi= \big[ t_{A,\kk \alpha}, t_{B,\kk \alpha}, t_{A,-\kk \alpha}^{\dagger}, t_{B,-\kk \alpha}^{\dagger}  \big]^{T}$, $\alpha=x,y,z$ is triplon flavor, A and B the two sublattices, the matrix $\mathcal{M}_{\kk,\alpha} = \mathbbm{1} \otimes h_{1,\kk\alpha} + \sigma_{1} \otimes h_{2,\kk\alpha}$ with
\begin{equation}
\label{eq:hh1}
h_{1,\kk\alpha} = \jp \mathbbm{1} + h_{2,\kk\alpha} \,, ~~~~
h_{2,\kk\alpha} =
\begin{bmatrix}
0 & \kappa_{\alpha} \\
\kappa_{\alpha}^{*} & 0
\end{bmatrix} \,.
\end{equation}
The parameter $\kappa_{\alpha}$ is defined as follows:
\begin{align}
\label{eq:kap}
\kappa_{\alpha} &= - \frac{K_{L}^{\alpha}}{2} e^{\ii \kk \cdot (\vec{L}_{1,\alpha} + \vec{L}_{2,\alpha})/2} \,, ~~~~~ \text{with} \nonumber \\
K_{L}^{\alpha} &= K^{\alpha} \cos\bigg[ \frac{\kk \cdot (\vec{L}_{1,\alpha} - \vec{L}_{2,\alpha})}{2} \bigg] \,,
\end{align}
where $\vec{L}_{1,\alpha} = \delta_{\alpha,x} \vec{a}_{1} + \delta_{\alpha,y} \vec{a}_{2}$ and $\vec{L}_{2,\alpha} = \delta_{\mu,\alpha} \vec{a}_{1} + \delta_{\nu,\alpha} \vec{a}_{2}$,
and $\vec{a}_{1,2}=\lbrace \pm \hat{x}/2,\sqrt{3}\hat{y}/2 \rbrace$ are the basis vectors of the triangular Bravais lattice.

At the level of this harmonic approximation, the triplon dispersion is simply given by the non-negative eigenvalues of the non-Hermitian matrix $\Sigma \mathcal{M}_{\kk,\alpha}$, where $\Sigma=\sigma_{3} \otimes \mathbbm{1}$ with
$\sigma_{3}$ being the Pauli matrix. Since $[h_{1,\kk\alpha}, h_{2,\kk\alpha}]=0$, the eigenvalues of $\Sigma \mathcal{M}_{\kk,\alpha}$ are straightforward to obtain and we thus have the
following triplon dispersion:
\begin{equation}
\label{eq:wk_bo}
\omega^{\alpha}_{A,B} = \sqrt{\jp \left( \jp \pm |K_{L}^{\alpha}| \right)} \,.
\end{equation}

For the AA stacking, i.e. $\mu\nu\delta \rightarrow xyz$, $\vec{L}_{1,\alpha}=\vec{L}_{2,\alpha}$ and so all the three triplons are dispersionless:
$\omega^{\alpha}_{A,B} = \sqrt{\jp \left( \jp \pm |K^{\alpha}| \right)}$. Within the harmonic approximation, here each triplon flavor is restricted to only one type of bond and
hence can not disperse. Actually this fact remains true even upon inclusion of the quartic terms. However, sixth order terms in triplons might add some dispersion.
At the harmonic level, the triplon gap closes for $\jp=\max(K^{x},K^{y},K^{z})$ at all points in the Brillouin zone.

For AB stacking ($\mu\nu\delta \rightarrow yzx$), the triplons are not restricted to a specific bond and can move along zigzag chains formed by bonds with same flavor from the two layers. For instance, the $t_y$ mode can move along zigzag chains formed by the $K^y$-bonds in layer-$1$ and layer-$2$. Thus the triplons have an effective one-dimensional dispersion, given by
\begin{align}
\label{eq:wx_AB_bo}
\omega^{x}_{A,B} &= \sqrt{\jp \left( \jp \pm \bigg|K^{x} \cos\left(\frac{K^x + \sqrt{3} K^y}{4}\right)\bigg| \right)} \,, \\
\label{eq:wy_AB_bo}
\omega^{y}_{A,B} &= \sqrt{\jp \left( \jp \pm \bigg|K^{y} \cos\left(\frac{K^x}{2}\right)\bigg| \right)} \,,  \\
\label{eq:wz_AB_bo}
\omega^{z}_{A,B} &= \sqrt{\jp \left( \jp \pm \bigg|K^{z} \cos\left(\frac{K^x - \sqrt{3} K^y}{4}\right)\bigg| \right)} \,.
\end{align}
The minima of the respective dispersion is along a line passing through the $\Gamma$ point. Thus the triplon gap will close at $\jp=\max(K^{x},K^{y},K^{z})$
along a line in the Brillioun zone connecting M-points on the opposite edges. Note that such a feature also arises in the bilayer Kitaev model on a triangular lattice.

In the case of AC stacking, i.e. $\mu\nu\delta \rightarrow xzy$, the $t_x$ triplon is confined to the $K^x$ bond only and hence it is dispersionless:
$\omega^{x}_{A,B} = \sqrt{\jp \left( \jp \pm |K^{x}| \right)}$. On the other hand, the remaining two
flavors of triplons move along zigzag chains and are degenerate if $K^{y}=K^{z}$. Therefore, in this case, depending on which triplon gap closes first, there will be gap closing either along a line or
in the entire Brillioun zone. For completeness, we quote the $t_{y,z}$ dispersion here:
\begin{equation}
\label{eq:w_AC_bo}
\omega^{y,z}_{A,B} = \sqrt{\jp \left( \jp \pm \bigg|K^{y,z} \cos\left(\frac{K^x - \sqrt{3} K^y}{4}\right)\bigg| \right)} \,.
\end{equation}
The overall dispersion results are perfectly consistent with that obtained from the dimer series expansions in Sec.~\ref{sec:series_expansion_dimer}. Apparently, the 1D dispersions reflect the approach to the MAC phases discussed in the main text.

Extending the bond-operator treatment beyond the harmonic level is possible \cite{kotov,larged_para,larged_af} but beyond the scope of the present work. We expect that the properties of \pddim{} are equally well captured by the dimer series.



\begin{thebibliography}{99}
\bibitem{fn:symmop}
The symmetry operation can be defined as follows: Define a zigzag string through the lattice (say, formed by $x$-$y$-bonds), and perform  $C^\ast_z$, $C^\ast_x$, $C^\ast_y$, $\mathds{1}$ on four successive sites, and repeat. Cover the lattice with these four operations by applying them to every opposite site on each hexagon.
\bibitem{fn:params}
Shown here are all parameters that are non-vanishing in at least one phase.
\bibitem{fn:temp}
For numerical reasons, we use a small finite temperature of $T=10^{-4}$ and we have checked that our results are consistent for $T\to 0$.
\end{thebibliography}
\end{document}